\documentclass[acmlarge]{acmart} 

\setcopyright{none}

\acmJournal{IMWUT}
\acmYear{2027}
\acmVolume{0}
\acmNumber{0}
\acmArticle{0}
\acmMonth{0}

\usepackage{amsmath,amssymb,amsfonts}
\usepackage{graphicx}
\usepackage{textcomp}
\usepackage{booktabs}
\usepackage{multirow}
\usepackage{makecell}
\usepackage{pifont}
\usepackage{algorithm}
\usepackage{algpseudocode}

\newcommand{\cmark}{\ding{51}}%
\newcommand{\xmark}{\ding{55}}%

\usepackage{xcolor}

\usepackage{subcaption}

\begin{document}

\title{Towards Trustworthy Mobile Speech Interaction: Ensuring Voice Authenticity and Transmission Integrity}

\author{Guodong Liu}
\orcid{0009-0004-2463-9586}
\affiliation{%
  \institution{Nanjing University}
  \city{Nanjing}
  \state{Jiangsu}
  \country{China}
}
\email{EricLiu@smail.nju.edu.cn}

\author{Lei Wang}
\authornote{Corresponding authors.}
\orcid{0000-0002-0091-0931}
\affiliation{%
  \institution{Soochow University}
  \city{Suzhou}
  \state{Jiangsu}
  \country{China}
}
\email{wanglei@suda.edu.cn}

\author{Dongxu Liu}
\orcid{0009-0008-8305-7198}
\affiliation{%
  \institution{Nanjing University}
  \city{Nanjing}
  \state{Jiangsu}
  \country{China}
}
\email{dongxuliu@smail.nju.edu.cn}

\author{Qianran Qiao}
\orcid{0009-0009-4767-5922}
\affiliation{%
  \institution{Nanjing University}
  \city{Nanjing}
  \state{Jiangsu}
  \country{China}
}
\email{522025330078@smail.nju.edu.cn}

\author{Haipeng Dai}
\authornotemark[1]
\orcid{0000-0003-0545-8187}
\affiliation{%
  \institution{Nanjing University}
  \city{Nanjing}
  \state{Jiangsu}
  \country{China}
}
\email{haipengdai@nju.edu.cn}

\author{He Huang}
\orcid{0000-0003-2768-6607}
\affiliation{%
  \institution{Soochow University}
  \city{Suzhou}
  \state{Jiangsu}
  \country{China}
}
\email{huangh@suda.edu.cn}

\author{Guihai Chen}
\orcid{0000-0002-6934-1685}
\affiliation{%
  \institution{Nanjing University}
  \city{Nanjing}
  \state{Jiangsu}
  \country{China}
}
\email{gchen@nju.edu.cn}

\begin{abstract} 
The rapid progress of generative speech models has made high-fidelity voice cloning and editing cheap, fast, and widely accessible, weakening the security of voiceprint-based authentication in mobile speech interactions.
Existing speech watermarking techniques focus on offline protection or watermarking synthetic speech, and thus cannot verify whether a live utterance originates from the genuine speaker at capture time.
We present \textit{MoBiMark}, a real-time \textbf{mo}bile \textbf{bi}ologic speech water\textbf{mark}ing system that binds ultrasonic articulatory signatures to speech content and embeds a dynamic watermark during capture.
MoBiMark emits a multi-frequency ultrasonic probe on commodity smartphones, extracts echo features from the vocal tract, and fuses them with a content-dependent speech embedding to produce a dual-layer dynamic watermark combining a persistent user identifier (UID) and a per-utterance dynamic speech identifier (SID), enabling both source and propagation-time attack detection.
A jointly trained embedder and detector injects the watermark while preserving perceptual quality, and system-level optimization with INT8 quantization enables lightweight on-device deployment.
Evaluated on 50 users across four smartphone models, 
MoBiMark achieves an average 99.79\% accuracy in source and propagation attacks, with negligible impact on speech quality and intelligibility.
\end{abstract}



\maketitle

\section{Introduction}\label{sec:intro}

Voice interaction has become a dominant interface for mobile assistants, smart speakers, and multimodal applications, yet the security of voice-based systems is increasingly fragile.
Contemporary artificial intelligence (AI) generative models can clone a target speaker's timbre from a short enrollment sample and synthesize fluent speech at scale, rendering voiceprint verification vulnerable to spoofing and content manipulation.
Many voice-interactive intelligent agents are now capable of directly manipulating applications and accessing private data on users' smart devices, yet they still rely on traditional voiceprint-based authentication mechanisms, posing significant privacy and security risks.

Large-scale anti-spoofing experiments demonstrate that even strong automatic speaker verification systems remain exposed to replay, conversion, and synthesis attacks across logical and physical access settings~\cite{Wang2020ASVspoof, ASVspoof2021, RawNet2, AASIST2021}.
Modern audio models built on neural architectures produce high-quality speech with low latency, while neural vocoders and end-to-end systems push naturalness closer to human recordings~\cite{WaveNet2016, Tacotron2, FastSpeech2, HifiGAN2020, VITS2021}.
\begin{figure}[h]
  \centering
  \includegraphics[width=0.9\linewidth]{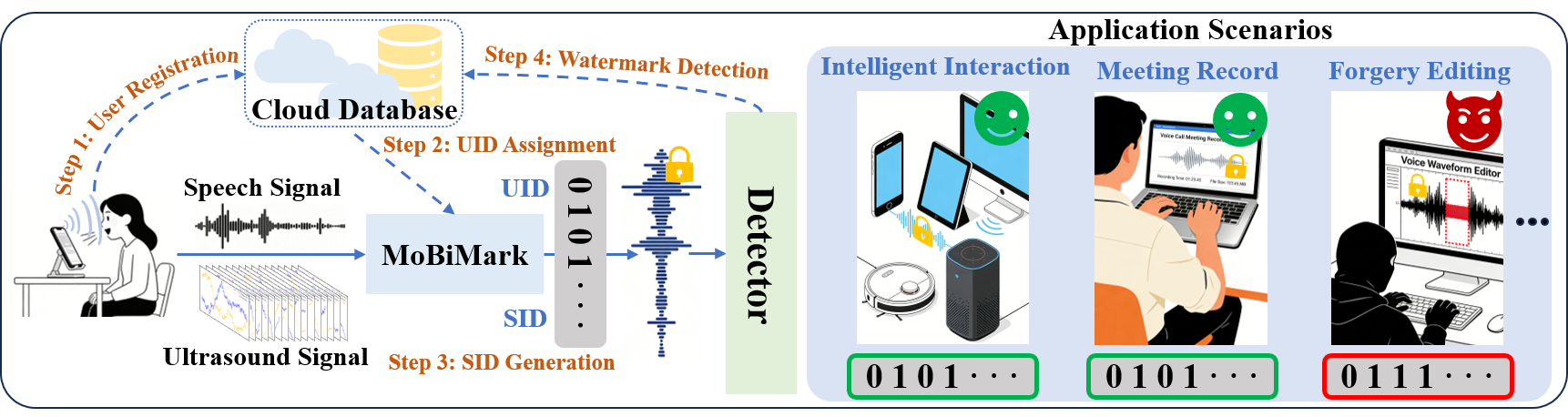}
  \caption{MoBiMark for speech security in different scenarios.}
  \label{fig: system overview}
  \vspace{-0.2in}
\end{figure}
Voice conversion methods enable style transfer and timbre conversion with limited training data~\cite{AutoVC2019, StarGANVC2}, and large-model-based one-shot inference methods can forge arbitrary speech from a single target sample~\cite{liu2024zero, wang2026onevoice, ding2025kimi}.
These advances drastically reduce the cost of producing convincing spoofed speech, creating a strong incentive for online real-time defenses that bind authenticity to the physical speaker.

Digital speech watermarking has emerged as a promising mitigation, with decades of work on imperceptible and robust audio watermarking~\cite{Swanson1998Audio, Hua2016Review, AlHaj2014Audio, cox2002digital}.
Recent learning-based watermarking frameworks improve robustness under compression and filtering~\cite{Pavlovic2022DNN, fu2024watermarking, chen2023wavmark, roman2024proactive, yao2025speechverifier}, but they still lack broad applicability.
In the face of advanced voice cloning, a watermark can be removed, replayed, or replaced by attackers.
Existing methods predominantly operate offline, focusing on protecting user-generated content after creation or verifying AI-generated output.
As AI-powered speech forgery capabilities grow increasingly sophisticated, applying watermarks to protect speech at the moment of utterance emerges as a more logical and urgent approach.
There is currently no work that focuses on using digital watermarking to protect user speech online in real time.
A watermarking system operating on a device still cannot verify that the input speech originates from a real user, which remains one of the key factors limiting wider adoption.

We observe two core limitations of existing speech watermarking systems.
First, content-only watermarks cannot verify the physical source: attackers can remove or regenerate watermarks using private models, or replay and re-edit captured speech to bypass detection.
Second, most designs use static watermarks and lack reliable biometric binding in online settings, where registration requirements and device-specific keys limit deployability.
A practical system must bind a watermark to the user's biometric characteristics during capture, operate without specialized hardware, and remain robust under common signal processing attacks.

Therefore, this paper presents \textit{MoBiMark}, an online real-time ultrasonic speech biometric watermarking system that binds vocal-tract motion to speech and embeds a dynamic watermark into every utterance at capture time.
As shown in Fig.~\ref{fig: system overview}, MoBiMark emits an inaudible multi-frequency ultrasonic signal via the phone earpiece and captures its reflections through the built-in microphone.
The reflected echoes encode signatures of articulator motion, which are fused with speech content to produce a biometric-dependent dual-layer watermark.
The watermark contains a stable UID and a content-dependent dynamic SID to enable both source and propagation attack detection.
The system can be practically deployed on users' smartphones, enabling real-time watermark integration at the moment of utterance.

Although our system shows great promise, several key challenges remain to be addressed:
%
First, how to effectively perform feature transformation and watermark encoding to ensure user speech protection throughout propagation. 
We propose a dual-layer dynamic watermark encoding method combining a system-assigned static UID with a speech-correlated dynamic SID.
This ensures that tampered or forged speech can be reliably detected. By integrating the source user’s biometric information and security mechanisms, a closed-loop security framework is formed from the source to the propagation process.  
Second, how to effectively embed the generated watermark into the original speech and achieve reliable detection, ensuring that the watermark remains detectable and supports accurate discrimination after embedding. 
In the embedding scenario, we enhance watermark robustness against benign modifications by adjusting the speech frame ordering strategy. Furthermore, through security-oriented analysis, we incorporate an external stabilizer to perform denoising and enhancement on both the encoded speech information and the input at the detection end, thereby ensuring the stability of the generated speech features and the watermark.  
Finally, how to build a universal and high-accuracy system on real-world mobile smartphones, enabling online real-time use on resource-constrained edge devices. 
Benefit to the widespread availability of microphone hardware required for ultrasonic signals, our system inherently possesses hardware universality. Combined with our carefully designed lightweight model, which employs efficient modules and incorporates quantization to maximally compress the original system while preserving accuracy, we successfully deploy the system on mobile devices, achieving a truly universal, secure and accurate solution.

In summary, our contributions are:

\begin{itemize}
\item To the best of our knowledge, MoBiMark is the first real-time online speech watermarking system that combines ultrasonic biometric signals with speech for proactive speech interaction security. Due to the widespread availability of microphone sensors supporting ultrasonic signals across various edge devices, MoBiMark can seamlessly integrate with a wide range of existing smart mobile devices, enabling speech security protection in numerous application scenarios, serving as a superior alternative to voiceprint-based methods.
%
\item We have carefully addressed three key challenges to bridge the gap between fundamental insights and practical application. These challenges span from source-level fusion to cloud-based detection, and finally to real-world deployment scenarios, thereby forming a closed-loop security framework for speech from source to propagation, as well as a practical realization from theoretical design to actual implementation. This ensures the effectiveness of the system in real-world settings such as on smartphones and mobile devices.
\item We deploy MoBiMark on four smartphone models and validate it using a dataset from 50 volunteers (25 hours of positive samples, 290+ hours of negative attack samples). Through carefully designed spoofing and editing attacks, the effectiveness of our dual-layer dynamic encoding combined with biometric speech watermarking is thoroughly verified.
Extensive experiments demonstrate average attack detection accuracy of 99.79\% and negligible perceptual impact, bridging the gap between research on online biometric
speech watermarking and practical deployment on edge mobile devices, offering a new means for protecting users’ online speech.
\end{itemize}

\section{Related Work}
\label{sec:related}

\subsection{Audio Watermarking}

Audio watermarking has a long history, with classic signal processing approaches embedding imperceptible information in time or transform domains~\cite{Swanson1998Audio, Hua2016Review, AlHaj2014Audio, Bassia2001, LieChang2006}.
Robustness and imperceptibility are typically achieved through psychoacoustic masking and transform-domain embedding, but these methods are often offline.
Recent advances include hybrid transform algorithms and learning-based embedder-detector architectures that improve robustness against common signal processing attacks~\cite{Pavlovic2022DNN,roman2024proactive,yao2025speechverifier}.
Speech-specific systems extend these ideas to linguistic content and speaker identity~\cite{Mawalim2021McAdams}, and timbre watermarking has been proposed to detect voice cloning attacks~\cite{Liu2024Timbre}.
Recent AI-generated content watermarking emphasizes model-side identifiers during synthesis, valuable for controlled pipelines but not for human speech captured in the wild.
MoBiMark differs by binding watermarks to biometric signatures at capture time with dynamic content-dependent encoding.

\subsection{Audio Forgery Detection}

The quality of spoofed speech is tightly linked to progress in speech synthesis and voice conversion~\cite{WaveNet2016, Tacotron2, FastSpeech2, HifiGAN2020, VITS2021, AutoVC2019, StarGANVC2}.
Anti-spoofing detectors face persistent generalization challenges~\cite{Wang2020ASVspoof, ASVspoof2021}, and models such as RawNet2 and AASIST, while strong on benchmarks, remain vulnerable to domain shift and provide no content integrity guarantees~\cite{RawNet2, AASIST2021}.
Defenses typically focus on detecting artifacts in frequency or phase statistics, but these signatures quickly become obsolete as synthesis models improve. 
Furthermore, detection-only systems face an asymmetry problem: false negatives may allow high-impact attacks, while false positives disrupt legitimate interactions.
MoBiMark complements anti-spoofing by integrating biometric watermarking into the capture pipeline, creating an auditable authenticity trail for downstream forensics.

\subsection{Acoustic and Ultrasound Sensing}

Ultrasound and active acoustic sensing have enabled contactless interaction and authentication on commodity devices.
EchoWhisper uses ultrasonic echoes to infer silent speech on smartphones~\cite{Gao2020EchoWhisper}.
VocalLock leverages acoustic sensing of the vocal tract for passphrase-independent authentication~\cite{Lu2020VocalLock}.
Speaker-Sonar employs inaudible sonar to verify voice command liveness against remote attackers~\cite{Lee2020SpeakerSonar}.
TouchPrint demonstrates active acoustic sensing for geometric biometric authentication~\cite{Chen2020TouchPrint}.
Compared with camera-based biometrics, ultrasound sensing is insensitive to lighting, operates unobtrusively with existing hardware, and captures subtle articulatory dynamics difficult to spoof remotely.
MoBiMark builds on these insights by fusing ultrasonic biometric features with speech to produce a propagation-protecting watermark, forming an end-to-end security framework.

\begin{table*}[t]
  \centering
  \caption{Comparison between MoBiMark and other methods.}
  \scalebox{0.82}{
  \begin{tabular}{c c c c c c c}
    \toprule
    Method & \makecell{Protects \\Speaker ID} & \makecell{Source-side \\Liveness} & \makecell{Protects \\Propagated Audio} & \makecell{Dynamic \\Speech Binding} & \makecell{Resistant to \\Speech Editing} & \makecell{Resistant to \\Replay / Clone} \\
    \midrule
    \makecell{Speaker Verification (Voiceprint)\\~\cite{shang2010score,wang2011channel,wu2015spoofing,kinnunen2017asvspoof}} & \cmark & \xmark & \xmark & \xmark & \xmark & Limited \\
    \makecell{Voice Liveness Detection\\~\cite{schroff2015facenet, yan2026comprehensive, abdulkareem2025hardware, shuwandy2025sensor}} & \cmark & \cmark & \xmark & \xmark & \xmark & Partial \\
    \makecell{Audio Watermarking\\~\cite{uddin2024audio,roman2024proactive,yao2025speechverifier,yao2025yours,chen2023wavmark}} & \xmark & \xmark & \cmark & Partial & Limited & \xmark \\
    \textbf{MoBiMark} & \cmark & \cmark & \cmark & \cmark & \cmark & Strong \\
    \bottomrule
  \end{tabular}}
\label{tab:compare}
  \vspace{-0.2in}
\end{table*}

\subsection{MoBiMark Difference}

%
Our watermarking system prevents source forgery through biometric authentication at the point of capture and protects speech during transmission using dynamic watermarking together with a corresponding detection mechanism. As a result, only speech carrying the correct watermark is accepted as authentic.
As shown in Table~\ref{tab:compare}, compared to conventional speaker verification (voiceprint) schemes~\cite{shang2010score,wang2011channel,wu2015spoofing,kinnunen2017asvspoof}, our method encompasses both the source and propagation stages, actively embedding a biometric watermark at capture time.
Compared to voice liveness detection schemes~\cite{schroff2015facenet, yan2026comprehensive, abdulkareem2025hardware, shuwandy2025sensor}, MoBiMark overcomes the limitation of verifying identity only at acquisition time by embedding dynamic speech-related watermarks that enable authenticity traceability throughout transmission, storage, and redistribution, while leveraging large-scale pre-trained speech model generalization to achieve cross-user capability without capturing a range of users' data for retraining.
Compared to traditional audio watermarking~\cite{uddin2024audio,roman2024proactive,yao2025speechverifier,yao2025yours,chen2023wavmark}, MoBiMark uniquely binds source-side liveness information to propagated speech, addressing the fundamental gap that existing watermarks cannot verify whether input speech originates from a real user. In summary, unlike all previous methods, we propose an integrated speech biometric watermarking system that protects all speech from the target human, from source to subsequent dissemination.

\section{Threat Model}
\label{sec:threat}

\subsection{Security Goals and Trust Assumptions}

MoBiMark aims to preserve speech authenticity throughout acquisition, transmission, and subsequent propagation by providing two security guarantees: (1) \textit{source authenticity}, ensuring that speech originates from a legitimate enrolled speaker, and (2) \textit{propagation integrity}, ensuring that propagated speech remains consistent with the originally generated content.

We assume that the user's smartphone, including the microphone, ultrasonic sensing module, operating system, and local watermark generation pipeline, operates correctly and is not compromised. Sensor measurements are acquired through the device's normal sensing interfaces. In particular, the adversary cannot directly tamper with sensor hardware or drivers, bypass the sensing pipeline to inject fabricated digital measurements, or otherwise manipulate the trusted acquisition process.
These assumptions define the trusted hardware and sensing boundary of MoBiMark. Attacks that compromise this boundary are outside the scope of our watermarking system. 

The ultrasonic evidence used by MoBiMark is generated by the smartphone's own speaker and captured by its microphone through the normal acoustic propagation path. The acquisition and binding of speaker-related biometric characteristics and the corresponding ultrasonic measurements are performed within the trusted sensing pipeline. We assume that the raw ultrasonic measurements and the internal correspondence between the target speaker's biometric characteristics and the captured ultrasonic responses are protected from direct adversarial access. In particular, the adversary cannot obtain target-specific paired speech-ultrasound observations sufficient to reconstruct the correspondence exploited by MoBiMark. This assumption concerns the protection of the trusted acquisition and binding process, rather than the physical impossibility of external ultrasonic interference.

The cloud verification service is trusted and follows the prescribed protocol, while communication among trusted components is protected using standard transport-layer security mechanisms. We further assume that user enrollment is performed in a trusted environment, consistent with existing biometric authentication systems such as Face ID and fingerprint authentication. Since the security of MoBiMark fundamentally depends on the correctness of the enrolled identity, impersonation during enrollment compromises the root of trust rather than exploiting a vulnerability in MoBiMark. Moreover, attacks against MoBiMark are meaningful only after a user has enrolled and enabled watermarking. We therefore consider enrollment attacks, as well as full compromise of trusted hardware or software, outside our threat model.

\subsection{Attack Model}



We consider a strong adversary who knows the general working principle the MoBiMark architecture. The adversary may obtain previously published speech recordings, corresponding watermarked speech, and access to the static detector implementation. Notably, the static detector differs from our cloud-based dynamic detector in that it can only detect the statically assigned watermark key, consistent with watermark detection schemes in most prior studies. In contrast, detection of the dynamic watermark is performed exclusively within our trusted cloud service, and the dynamic detection mechanism remains undisclosed to the public.

The adversary can perform replay attacks, voice cloning, speech conversion, speech editing, watermark reuse, and other digital manipulations in an attempt to bypass the verification process.
We further consider adaptive adversaries who are aware of the speech-ultrasound consistency constraints exploited by MoBiMark. Such adversaries may attempt to jointly construct synchronized speech and ultrasonic signals, replay forged speech while presenting artificial ultrasonic signals, or reuse previously observed content to maximize their acceptance probability. However, these attacks are constrained by the trusted acquisition and binding assumptions described above.

The adversary cannot directly inject fabricated measurements into the smartphone's sensor data stream or access the internal correspondence between the target speaker's biometric characteristics and the captured ultrasonic responses. Consequently, attacks that require eavesdropping on the target speaker's ultrasonic responses or collecting sufficient target-specific paired speech-ultrasound observations to train a model that reproduces this correspondence are outside our threat model. Similarly, the adversary is assumed not to have access to synchronized speech-ultrasound pairs collected during legitimate watermark generation when such access would expose the protected correspondence. These restrictions do not imply that external ultrasonic signals are physically incapable of reaching the microphone; rather, they reflect the protection of the trusted acquisition and binding process.

Accordingly, MoBiMark focuses on detecting forged or manipulated speech and preserving the association between legitimately captured speaker-related biometric signals and the resulting watermarked speech throughout subsequent propagation. Direct physical attacks on trusted sensors, malicious modification of sensor drivers or firmware, privileged compromise of the operating system, and other attacks that violate the trusted sensing boundary are outside the scope of this work.
In addition, strengthening hardware-level encryption and data-flow integrity protection is a promising direction for further hardening the system against unauthorized access to or injection of internal hardware data.


\begin{figure}[h]
  \centering
  \includegraphics[width=0.7\linewidth]{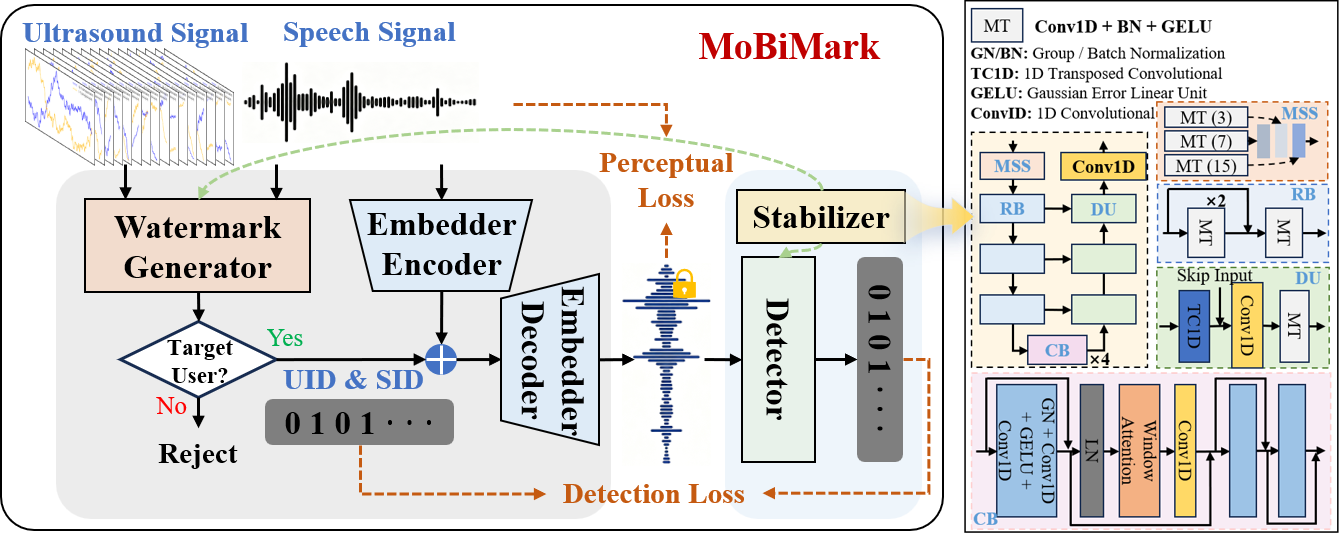}
  \caption{Overall architecture of MoBiMark.}
  \label{fig: MoBiMark overview}
  \vspace{-0.2in}
\end{figure}

\section{System Design} \label{sec:design}
MoBiMark is deployed on the client side of users' mobile devices.
The overall workflow, as illustrated in Fig.~\ref{fig: system overview}, involves the following components.

\noindent\textbf{1. User Registration:}
Before using the system, a user registers their voice and ultrasonic data with a single arbitrary sample, sufficient due to the pre-trained encoder's generalization capability. 
Specifically, the registered samples are converted into compressed feature vectors, stripped of the user's biometric information, and then uploaded to a trusted cloud for subsequent watermarking and detection.

\noindent\textbf{2. UID Assignment:}
After registration, the system automatically assigns a static 64-bit UID code, which will be combined with the subsequent dynamic SID.

\noindent\textbf{3. SID Generation:}
During actual voice interaction, MoBiMark automatically and dynamically generates a content-dependent 192-bit SID for each utterance.
The encoded information is strongly correlated with the speech features; if the speech is modified, the encoded value changes accordingly, enabling effective real-time dynamic protection.

\noindent\textbf{4. Watermark Detection:}
For simple verification scenarios, the detector can use the UID locally, similar to mainstream static watermarking schemes.
For forged speech that has been meticulously modified or edited by an attacker, the speech only needs to be uploaded to the cloud, where the system regenerates the corresponding watermark information based on the current speech and compares it with the embedded watermark, returning the comparison result.
%

As shown on the left of Fig.~\ref{fig: MoBiMark overview}, the architecture consists of the Watermark Generator, the Embedder Encoder/Decoder, the Detector and Stabilizer modules. Specifically, the concurrently captured speech and ultrasound signals are fed into MoBiMark. They are first examined and compared by the Watermark Generator module; if a live user is verified, a dual-layer dynamic watermark code (UID + SID) is generated based on both signals and then combined with the intermediate speech features produced by the Embedder Encoder. The result is decoded by the Embedder Decoder into a watermarked speech signal that preserves the original speech quality. The subsequently deployed Detector extracts the embedded watermark information. The Stabilizer is applied within both the Watermark Generator and the Detector to maintain the stability of the watermark features.
Benefiting from the ubiquity of microphone sensors required for ultrasound on all mobile phones, MoBiMark exhibits a high degree of universality.

\begin{figure}[t]
  \centering
  \includegraphics[width=0.5\linewidth]{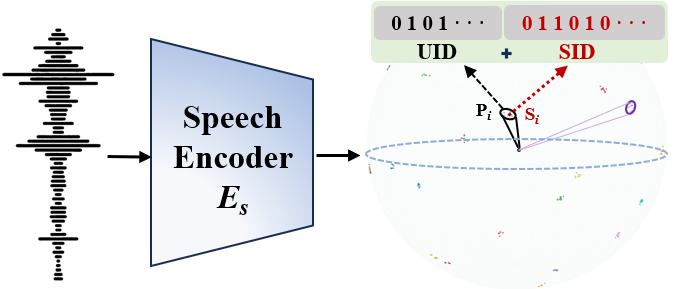}
  \caption{Speech feature clusters for different registered users.}
  \label{fig:cluster}
\end{figure}

\subsection{Watermark Generation}
\subsubsection{Speech Encoder}

The design goal of MoBiMark is a novel watermarking system capable of integrating with speech information while ensuring real-time security at the source. We abstract the most fundamental requirements of the system: it must be able to distinguish between a large number of newly registered different users, ensuring feature independence among them. Simultaneously, it should also differentiate between different speech utterances from the same user, in order to meet the subsequent demands of dynamic watermarks that are strongly correlated with speech content.
During training, the system utilizes speech data from known users to construct an effective feature space. This feature space must ensure that speech features from different users form distinct clusters, while also allowing differentiation among different speech features from the same user, as illustrated in the Fig.~\ref{fig:cluster}.

From the perspective of theory, the speech representation learning task addressed by this system can be modeled as a joint optimization problem of category discrimination and sample diversity.
Let the high-dimensional observed feature embedded by the model be $X \in \mathbb{R}^{D}$, and the category label be $Y$. The learning objective can be uniformly formulated as maximizing a weighted combination of the category mutual information and the within-class conditional entropy:

\begin{equation}
\max_\theta \ I_\theta(X;Y) + \lambda \cdot \mathbb{E}_{Y}[H_\theta(X|Y)]
\end{equation}

Here, $I_\theta(X; Y)$ represents the discriminative information associated with the category. Its maximization ensures significant distinguishability between different identity categories. Conversely, the conditional entropy $H_\theta(X|Y)$ measures the uncertainty and richness of the sample distribution within a given category. Its non-degeneracy constraint guarantees sufficient diversity among different samples belonging to the same class.

However, a feature space derived solely from this optimization objective is limited by our finite collection of sample data and cannot form a stable and effective representation. Consequently, it fails to generalize well when encountering new users.
We aim to train such an ultra-large-scale and stable vocal timbre feature space on a large number of different users, and ensure that newly registered users are effectively assigned to fixed regions within this feature space.

Inspired by speaker recognition tasks \cite{hannun2014deep, chan2016listen, wang2023cam++}, we observe that for such tasks, speech from different users forms distinct clusters in the feature space, with each cluster representing unique vocal timbre information. However, due to the leakage of semantic information in speech, different utterances from the same user do not completely converge to a single point; instead, they distribute at different locations within the cluster. While this is considered a drawback in speaker recognition tasks, it precisely aligns with the requirements of our task.
We leverage the encoder of the well-established CAM++ model \cite{wang2023cam++}, which is trained on a large-scale speaker dataset, as our Speech Encoder $E_s$. The CAM++ is trained on a dataset comprising over 200,000 user speech samples, enabling it to learn a broad range of vocal timbre information and achieve effective high-dimensional space embeddings for the vocal timbre of all users. Nevertheless, it is extremely lightweight and highly suitable for our system.
The 512-dimensional high-dimensional space output by CAM++ provides sufficient representational capacity for the aforementioned objective. The high dimensionality supports the formation of adequate inter-class informational separation, thereby increasing $I_\theta(X; Y)$. Simultaneously, it allows each class to occupy a non-degenerate support region on the high-dimensional manifold, thereby maintaining a relatively large $H_\theta(X|Y)$ and preventing excessive convergence of representations within the same class. Therefore, under this theoretic framework, the model achieves robust and highly discriminative speech representation learning by simultaneously enhancing inter-class separability and intra-class diversity within the high-dimensional space.

\begin{figure}[t]
  \centering
  \includegraphics[width=\linewidth]{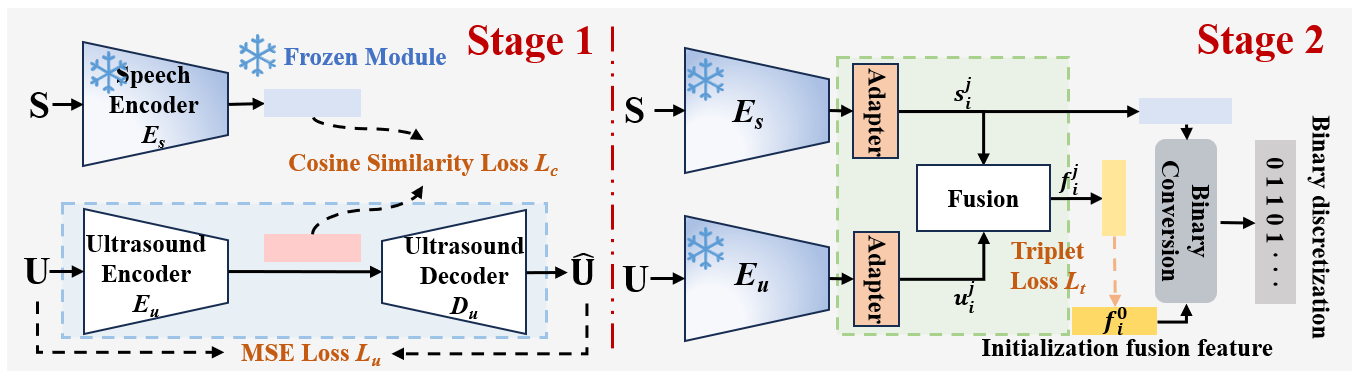}
  \caption{Overall architecture of the Watermark Generator.}
  \label{fig:wmg}
\end{figure}

\subsubsection{Ultrasound Encoder}

\begin{figure}[t]
  \centering
  \includegraphics[width=0.9\linewidth]{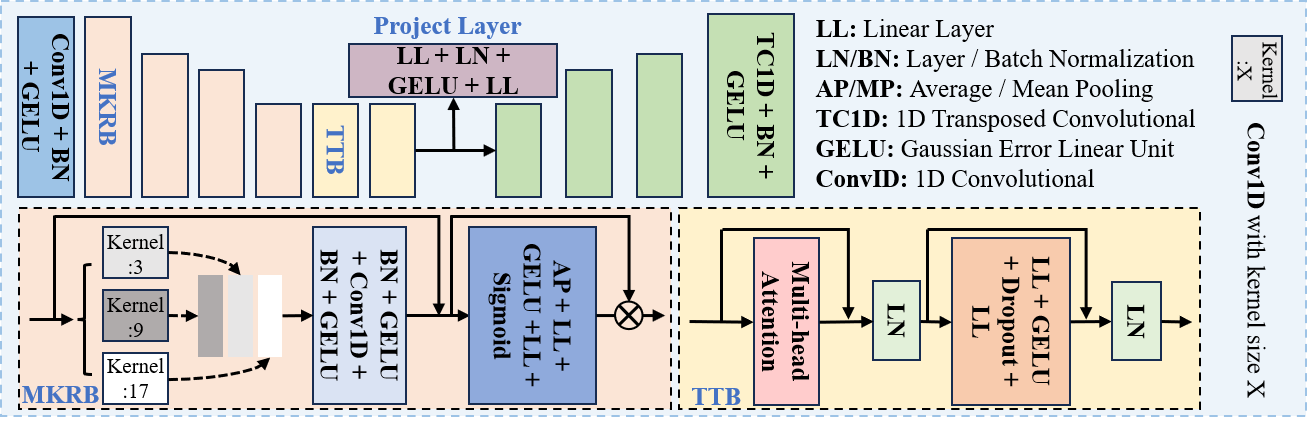}
  \caption{Architecture of ultrasound Encoder and Decoder.}
  \label{fig:wmg_s1}
  \vspace{-0.2in}
\end{figure}

We encode the speech modality using $E_s$, and simultaneously encode ultrasound signals with $E_u$, as shown in Fig.~\ref{fig:wmg_s1}.
%
To ensure the generalization ability of MoBiMark for newly registered users, it is necessary for $E_u$ to also exhibit generalization beyond the limited training samples.
If training is performed only on a limited number of data samples, the feature space learned by the model is confined to the regions corresponding to these samples.
When a new user registers, the feature space allocated to them tends to overlap with that of the training users, making effective differentiation impossible.
Therefore, we align ultrasound signal features and speech signal features in the high-dimensional feature space.
Since $E_s$ is trained on a very large-scale speech dataset, it can form an extremely large and stable feature space.
After aligning the speech features with the corresponding ultrasound signal features in this space, the ultrasound features are also able to learn an effective feature distribution within the corresponding feature space.
Compared to partitioning the feature space based on limited data, this approach substantially expands the scope of the feature space, ensuring that when a new user registers, the region assigned to their features is sufficiently distinct from those of other users.

We train $E_u$ using cosine similarity loss $\mathcal{L}_{\text{c}}$ with the output of $E_s$, combine with an MSE reconstruction loss $\mathcal{L}_{\text{u}}$ with decoder $D_u$ (the left in Fig.~\ref{fig:wmg}):
\begin{equation}
    \mathcal{L}_{\text{s1}} = \lambda_c \mathcal{L}_{\text{c}} + \lambda_u \mathcal{L}_{\text{u}}
\label{eq:stage1}
\end{equation}
with $\lambda_c=1.0$ and $\lambda_u=0.2$. 
We use $\mathcal{L}_{\text{c}}$ to align the ultrasound and speech features, which is designed to train $E_u$ to cluster samples from the same user together while separating those from different users, thereby achieving user discrimination. To ensure that samples within the same cluster remain distinguishable, a prerequisite for constructing the SID, the encoder output features are simultaneously passed through a weak decoder $D_u$ to reconstruct the original ultrasound information. This constitutes an adversarial training scheme: the weak decoder forces the model to preserve intra-class variations while still learning discriminative inter-user boundaries, so that distinct user clusters form without causing feature collapse within each class.

As shown in Fig.~\ref{fig:wmg_s1}, $E_u$ processes 32-channel raw waveform segments (shape [B, 32, 3000$\times T$]) adapted for ultrasound signals, where 3000 is the ultrasonic sampling rate and $T$ is the time length.
The encoder first performs local pre-encoding through a convolutional stem ($32\!\to\!96, k\!=\!9$), followed by four stacked multi-kernel residual blocks (MKRB) that employ strides of 5,5,4,3 to achieve a total temporal downsampling factor of $300\times$, compressing the time length from $3000\times T$ to $10\times T$.
Each MKRB consists of three parallel convolutional branches (kernel sizes of 3, 9, 17) to explicitly capture short-range, medium-range, and long-range waveform patterns. After branch fusion, the features undergo convolutional post-processing, followed by the addition of a $1 \times 1$ residual projection, and are then adaptively re-calibrated via a channel attention gating mechanism (compression ratio 8), achieving a balance between multi-scale modeling and channel selection.
On this basis, we cascade two temporal Transformer modules (TTB, embedding dimension 512, 8 attention heads, dropout 0.1) to further model global dependencies across segments.
After temporal average pooling, a two-layer projection head (Linear--LayerNorm--GELU--Linear) produces a 512-dimensional discriminative vector.
The decoder $D_u$ employs a four-stage deconvolutional pathway (strides 3,4,5,5) to progressively restore temporal resolution, using linear interpolation for length alignment when lengths are not evenly divisible. This design concurrently addresses identity discrimination capability and temporal structure fidelity.

\subsubsection{User Verification and Dynamic Watermark Generation}

Overall, our model training is divided into two stages, as shown in the Fig.~\ref{fig:wmg}. The first stage mainly includes the design and training of the aforementioned $E_s$ and $E_u$ modules, which are used to effectively encode speech and ultrasound signals into a feature space that allows for comparison and fusion. The features obtained from the two modalities are then used for subsequent user verification and dynamic watermark generation.  

\textbf{Verification:}
%
As shown on the right side of the Fig.~\ref{fig:wmg}, the output vectors of the two encoders are first transformed by the Adapter module to adapt to the encoding dimension required in later stages. The two vectors are then processed by the Fusion module, which outputs the fused feature vector $f_i^j$.  
To enable effective dimension conversion for new registered users and to assign the fused feature vector to an appropriate location in the fused feature space, we employ a triplet loss function $\mathcal{L}_{\text{t}}$ for metric learning. This loss function pulls features of the same user closer in the feature space, while pushing features of different users farther apart. The formula for this loss function is as follows:  
\begin{equation}
    \mathcal{L}_{\text{t}} = \max \left( \|\mathbf{x}_a - \mathbf{x}_p\|_2^2 - \|\mathbf{x}_a - \mathbf{x}_n\|_2^2 + \alpha, 0 \right)
\end{equation}
Where $\mathbf{x}_a$ is feature of the anchor point, $\mathbf{x}_p$ is positive sample feature same as user of anchor point, $\mathbf{x}_n$ is negative sample feature different from user of anchor point, and $\alpha$ is the margin, which controls the minimum distance between the positive and negative sample pairs.
Benefiting from the pre-trained modality encoders from the previous stage, the modules trained in this stage can easily cluster effective sample pairs of the two modalities from the same user, while effectively separating clusters from different users. This results in features that match the watermark encoding dimension, as well as corresponding fused features.  

\begin{figure}[t]
  \centering
  \includegraphics[width=0.9\linewidth]{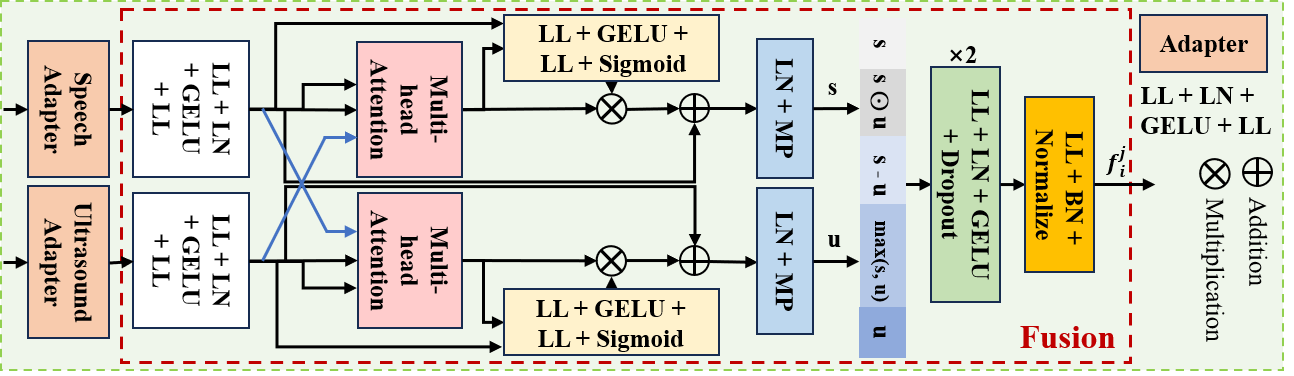}
  \caption{Architecture of Adapter and Fusion modules.}
  \label{fig:wmg_s2}
\end{figure}

Before fusion, we incorporate symmetric adapters for the ultrasound and audio modalities, respectively, in order to mitigate the statistical distribution gap between heterogeneous modalities and align the subsequent encoding dimensions. As shown in Fig. \ref{fig:wmg_s2}, each Adapter module employs a lightweight structure of Linear–LayerNorm–GELU–Linear, while preserving the original temporal length unchanged
Through this module, we can adaptively adjust the number of watermark encoding bits without affecting the training of other modules. We set the output dimension to 192, which is the same as the encoding dimension of the SID.

For the Fusion module, we adopt a bidirectional multi-head cross-attention mechanism, with an attention embedding dimension of 192 and 8 heads, enabling the ultrasound branch to query audio context and the audio branch to query ultrasound context in reverse, thereby explicitly modeling a bidirectional complementary relationship.
To suppress cross-modal noise propagation, a gating unit is applied after the attention outputs. The gating network takes the concatenated representation of the original features and the cross-attention features as input, with an input channel size of 384, and outputs gating coefficients of 192 dimensions. Residual connections and LayerNorm are further employed to enhance training stability.
In the global aggregation stage, the module pools the two sequences separately to obtain two 192‑dimensional global vectors ($s$ and $u$) and further constructs three types of relational features (multiplicative interaction $s \odot u$, difference response $s - u$, and dominant response $max\{s, u\}$), each also 192‑dimensional. The final concatenated fusion vector has a channel dimension of 960 (including $s$ and $u$). A metric head projects it as 960→384→192→192, and incorporates LayerNorm, BatchNorm, and Dropout layers.
This module ultimately outputs a fused embedding vector for the two modalities, with a dimension of 192, which is then used to match the final watermark encoding.

Using the well-trained features described above, we can obtain speech and ultrasound features of different users. It is only necessary to construct the corresponding logic to determine whether the input speech signal originates from the genuine target user. The specific algorithmic logic is shown in the Algorithm \ref{alg:verify}.

\begin{algorithm}[h]
\caption{Target User Verification}
\label{alg:verify}
\begin{algorithmic}[1]
\Require Speech $\mathbf{s}^j$, ultrasound $\mathbf{u}^j$, fused $\mathbf{f}^j$; enrolled features $\{(\mathbf{s}_i^0,\mathbf{u}_i^0,\mathbf{f}_i^0)\}_{i=1}^N$
\Require Thresholds $\gamma_s,\gamma_u,\gamma_f,\gamma_{\text{enroll}}$
\Ensure Verification result
\State Find best-matching user: $\text{best\_sim} \leftarrow \max_i \text{Sim}(\mathbf{s}^j, \mathbf{s}_i^0)$, $i \leftarrow \arg\max$
\If{$\text{best\_sim} < \gamma_{\text{enroll}}$} \Return (\texttt{False}, 0) \Comment{No match, enroll}
\EndIf
\If{$\text{best\_sim} < \gamma_s$} \Return (\texttt{False}, -1)
\EndIf
\State $Sim_u \leftarrow \text{Sim}(\mathbf{u}^j, \mathbf{u}_i^0)$, $Sim_f \leftarrow \text{Sim}(\mathbf{f}^j, \mathbf{f}_i^0)$
\If{$Sim_u \geq \gamma_u$ \textbf{and} $Sim_f \geq \gamma_f$} \Return (\texttt{True}, $i$)
\Else \Return (\texttt{False}, -1)
\EndIf
\end{algorithmic}
\end{algorithm}

The system first compares the input speech feature $\mathbf{s}^j$ with the enrolled speech features $\{\mathbf{s}_i^0\}_{i=1}^N$ to identify the best-matching user index $i$ and the corresponding similarity score $\text{best\_sim}$. If $\text{best\_sim} < \gamma_{\text{enroll}}$, the system identifies the input as belonging to an unregistered user and triggers the initial enrollment process, returning $(\texttt{False}, 0)$. Otherwise, if $\text{best\_sim} < \gamma_s$, the system rejects the verification and returns $(\texttt{False}, -1)$. For a candidate that passes speech verification, the system further computes the similarities $Sim_u$ and $Sim_f$ between the input ultrasound and fused features, $(\mathbf{u}^j, \mathbf{f}^j)$, and the corresponding enrolled features, $(\mathbf{u}_i^0, \mathbf{f}_i^0)$. The verification succeeds only if both $Sim_u \geq \gamma_u$ and $Sim_f \geq \gamma_f$, in which case the system returns $(\texttt{True}, i)$; otherwise, it rejects the input and returns $(\texttt{False}, -1)$. This hierarchical decision process uses speech similarity for candidate identification and preliminary verification, followed by joint ultrasound and fused-feature verification to strengthen authentication reliability. In our experiments, we set $\gamma_s = \gamma_u = \gamma_f = 0.5$ and $\gamma_{\text{enroll}} = 0.1$.
If all conditions are satisfied, the system authenticates the target user and embeds a dual-layer dynamic watermark into the speech. Otherwise, watermark embedding is rejected. Since downstream processing treats speech without a valid watermark, including speech carrying an incorrect watermark, as invalid, only authenticated speech with valid watermark information is accepted, ensuring source-level speech security.

\textbf{Watermark Generation:}
%
Beyond protecting speech at the source, it is essential to preserve its authenticity and integrity during transmission. Existing watermarking schemes often rely on static watermarks or lack real-time dynamic verification, limiting their effectiveness against replay and content-tampering attacks. To address these limitations, MoBiMark introduces a biologically grounded, dual-layer dynamic watermark comprising a static UID and a dynamic SID. The UID is a system-assigned 64-bit code that supports lightweight identity verification, while the 192-bit SID provides stronger protection for security-critical applications.

For each speech sample $\mathbf{S}_i^j$ and its synchronized ultrasound signal $\mathbf{U}_i^j$, pretrained encoders extract speech and ultrasound features, which are aligned through a dimension transformation module (DTM) and fused into a multimodal representation. The SID is then generated by combining the normalized difference between the current speech feature and the user's registered speech feature with the normalized registered fusion feature, followed by nonlinear activation and binarization:
\begin{equation}
    SID_i^j = \text{sign}\big(\tanh\big(\text{norm}(\hat{\mathbf{s}}_i^j - \mathbf{s}_i^0) \odot \text{norm}(\mathbf{f}_i^0)\big)\big)
\label{eq:sid}
\end{equation}

The final watermark is $\mathbf{B}_i^j=\operatorname{Concat}(UID_i,SID_i^j)$. By binding a persistent user identity to sample-specific speech and biometric features, the dual-layer design supports efficient identity verification while enabling dynamic content-integrity checks. Because the SID depends on the utterance and registered biometric information, semantic modifications (e.g., deleting ``not'') can be detected by checking the consistency between the received speech and its SID, while unauthorized speech cannot obtain a valid watermark through the source authentication process.

\subsection{Watermark Embedding and Detection}

After generating the dual-layer dynamic speech biometric watermark, it is embedded into the speech signal while ensuring \textit{imperceptibility} to human listeners and ASR systems, \textit{tamper sensitivity}, and \textit{robustness} against benign signal processing (e.g., resampling, filtering, and compression).

\subsubsection{Speech Stabilizer}


As shown on the right side of Fig.~\ref{fig: MoBiMark overview}, the Speech Stabilizer is shared by the watermark generation and detection stages to improve robustness against benign speech modifications, such as mild noise and down-sampling. During watermark generation, it denoises the input speech to obtain more stable speech and fusion features, improving subsequent cloud-based watermark detection. Importantly, denoising is applied only to the speech fed into the watermark generation module; the original speech remains unchanged to preserve its authenticity and contextual information.

During cloud-based detection, the stabilizer processes the potentially modified speech before regenerating the dynamic watermark for comparison with the watermark embedded in the speech. This design compensates for the sensitivity of the speech encoder $E_s$ to benign modifications, which enables dynamic watermark generation and helps defend against editing and forgery attacks. Meanwhile, joint anti-interference training allows the detector to directly recover watermark information despite such modifications. The stabilizer is optional: when tolerance to subsequent modifications is not required, any modification after watermark embedding can instead be treated as potential forgery, and the stabilizer can be omitted.

The stabilizer adopts a U-Net-like encoder-decoder architecture. A multi-scale convolutional stem (MSS; kernel sizes 3, 7, and 15) extracts local acoustic features, followed by three stride-2 convolutional layers with depthwise-separable residual blocks and increasing dilation rates. Four lightweight Conformer blocks form the bottleneck, combining feed-forward networks, windowed multi-head self-attention, and depthwise-separable convolution to capture contextual information with reduced sequence complexity from $O(T^2)$ to $O(Tw)$, where $T$ is the sequence length and $w$ is the attention window size. The decoder uses transposed convolutions and encoder skip connections to reconstruct the enhanced waveform through residual learning. Training minimizes a weighted combination of waveform MSE and Mel-spectrogram MSE, where the Mel spectrogram uses a 25-ms Hann window, a 10-ms frame shift, and 80 Mel filter banks:
\begin{equation}
    \mathcal{L}_{\text{stabilizer}}
    = \lambda_{mw}\mathcal{L}_{\text{mw}}
    + \lambda_{mm}\mathcal{L}_{\text{mm}},
\end{equation}
where $\lambda_{mw}$ and $\lambda_{mm}$ control the respective loss weights, and we set 10 and 1, respectively, by experimental training experience.

\subsubsection{Watermark Embedding}

\begin{figure}[t]
  \centering
  \includegraphics[width=\linewidth]{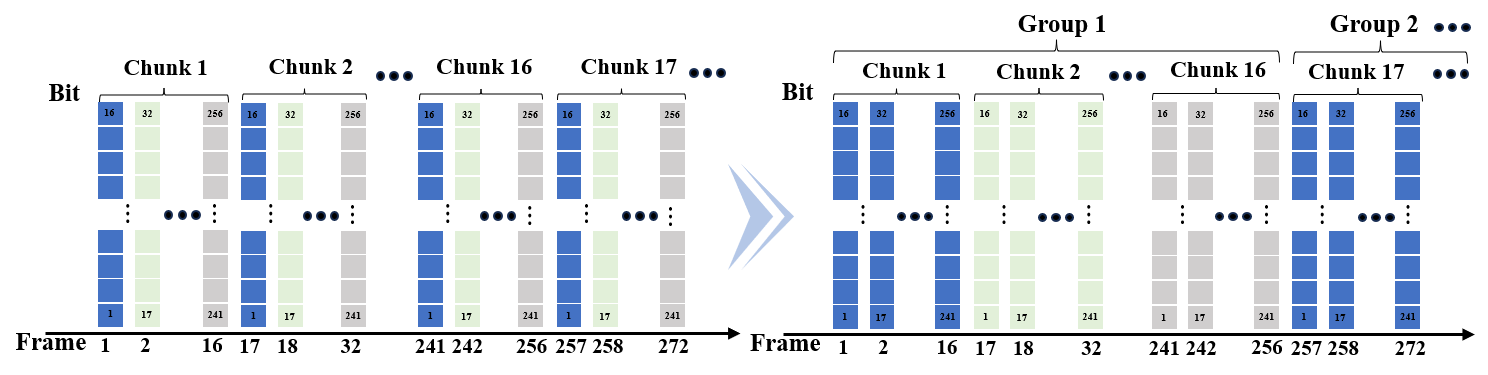}
  \caption{Chunk-based vs.\ group-based watermark framing.}
  \label{fig:group}
\end{figure}


We follow the design of AudioSeal \cite{roman2024proactive} and extend its original 16-bit watermark to a dynamic 256-bit watermark that can be embedded within a single utterance, drawing inspiration from SpeechVerifier \cite{yao2025speechverifier}. Specifically, we adopt a temporal framing strategy that divides the watermark into 16 frames, each encoding 16 bits. A spoken word typically lasts approximately 250 ms \cite{levelt1993speaking}, corresponding to around 4,000 audio samples at a sampling rate of 16 kHz. This provides the temporal context for embedding the complete watermark within an utterance.

As illustrated in the upper part of Fig.~\ref{fig:group}, a straightforward approach is to arrange the 16 watermark frames consecutively within each chunk, which we refer to as chunk-based framing. However, uniform downsampling may consistently remove the same relative frame positions across chunks, potentially causing systematic information loss and compromising watermark recovery. To mitigate this vulnerability, we propose group-based framing, illustrated in the lower part of Fig.~\ref{fig:group}. Specifically, each of 16 consecutive chunks carries one of the 16-bit watermark segments, repeated across 16 consecutive frames within that chunk. These 16 chunks form a group of 256 frames, across which the complete 256-bit watermark is distributed through temporal interleaving. By spreading watermark information across multiple chunks, this design is expected to improve tolerance to uniform downsampling and reduce the risk of systematic information loss caused by the removal of fixed relative frame positions. However, it does not guarantee the recovery of every watermark bit under all downsampling conditions.

Beyond the encoding strategy, the Perceptual Loss and Detection Loss play a critical role in balancing watermark imperceptibility, robustness to benign speech modifications, and sensitivity to malicious manipulation. We adapt the loss designs of AudioSeal and SpeechVerifier \cite{roman2024proactive,yao2025speechverifier} to jointly support these objectives.

\subsubsection{Watermark Detection}

The watermark detection component is primarily divided into edge-side and cloud-side detection, corresponding to the detection of static UIDs and dynamic SIDs, respectively.

\textbf{Edge Static Detection:}
Deployed on edge devices, this detection method resembles most existing watermark detection endpoints and is applicable only to simple or real-time, non-falsifiable security scenarios, such as in smart Internet of Things (IoT) device security detection. By verifying static UIDs, edge devices can determine whether the current speech originates from the target user. 
In practice, certain benign modifications may affect detection accuracy, leading to discrepancies between the extracted watermark code and the true code. Therefore, the Hamming distance \cite{hamming1950error} is employed to measure similarity by counting differing bit positions; a smaller value indicates higher similarity. The specific discrimination threshold can be set by the user.

\textbf{Cloud Dynamic Detection:}
However, in more complex scenarios requiring higher security, detection based solely on static UIDs is insufficient.
In high-security scenarios, compared to traditional watermarking systems, MoBiMark offers distinct advantages. Since the system ensures correctness from the source, it operates as a closed-loop system. For dynamic watermarking, the speech to be verified can be uploaded to the cloud for dynamic detection. The cloud system then returns a determination to the user regarding whether the speech originates from a genuine user.  
The principle of cloud detection involves generating corresponding watermark information from the speech to be verified through the Watermark Generator module. This generated watermark is then compared with the dynamic watermark code (SID) embedded in the speech. A match indicates authenticity; otherwise, the speech is considered inauthentic.  
In the cloud detection stage, the Stabilizer is utilized to recover speech that has undergone benign modifications, thereby facilitating more accurate discrimination of watermark information.

\begin{table}[h]
  \centering
  \caption{Module sizes before and after quantization.}
  \begin{tabular}{ccc}
    \toprule
    Module & FP32 (MB) & INT8 (MB) \\
    \midrule
    Watermark Generator & 81.37 & 24.97 \\
    Embedder & 56.17 & 21.57 \\
    Detector & 33.12 & 12.29 \\
    Stabilizer (Optional) & 84.00 & 23.88 \\
    \textbf{Total} & \textbf{170.66 (+84.00)} & \textbf{58.83 (+23.88)} \\
    \bottomrule
  \end{tabular}
\label{tab:quant}
\vspace{-0.1in}
\end{table}

\subsection{Lightweight Deployment}

Model quantization is an effective approach to edge deployment, reducing memory footprint and potentially accelerating inference by lowering the numerical precision of model weights and activations \cite{jacob2018quantization}. Compared with pruning and knowledge distillation, it can often be applied with limited architectural changes and lower retraining overhead. In this work, we quantize MoBiMark from FP32 to INT8 to improve deployment efficiency on resource-constrained smartphones. For affine quantization, a floating-point value \(x\) is mapped to an integer representation as \(x_q=\operatorname{clip}(\operatorname{round}(x/s)+z,q_{\min},q_{\max})\), and reconstructed as \(\hat{x}=s(x_q-z)\), where \(s\) and \(z\) denote the quantization scale and zero-point, respectively. We use symmetric quantization for weights and asymmetric quantization for activations to accommodate their different value distributions \cite{jacob2018quantization}.

To mitigate quantization-induced accuracy degradation, we adopt quantization-aware training (QAT), which simulates INT8 inference during training while retaining FP32 parameter optimization. Specifically, fake-quantization operators are applied to weights and activations, with the forward pass defined as \(\hat{x}=Q(x)=\operatorname{DeQuant}(\operatorname{Quant}(x))\). Since quantization is non-differentiable, the straight-through estimator (STE) approximates the gradient during backpropagation \cite{bengio2013estimating}. This allows the model to adapt to quantization errors before deployment. Layers unsupported by INT8 retain FP16 representations. As shown in Table~\ref{tab:quant}, the total model size decreases from 170.66 MB to 58.83 MB, a reduction of 65.53\%. All subsequent experiments use the resulting edge-deployed models, and we compare their performance with the original FP32 models to evaluate the impact of quantization.

\section{Experiments}
\label{sec:exp}

\subsection{Dataset and Experimental Setup} 

\begin{figure}[h]
  \centering
  \includegraphics[width=0.4\linewidth]{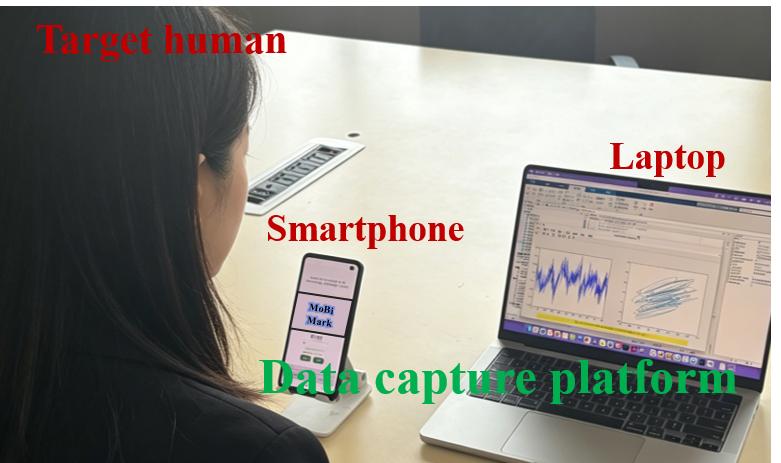}
  \hspace{0.04\linewidth}
  \includegraphics[width=0.44\linewidth]{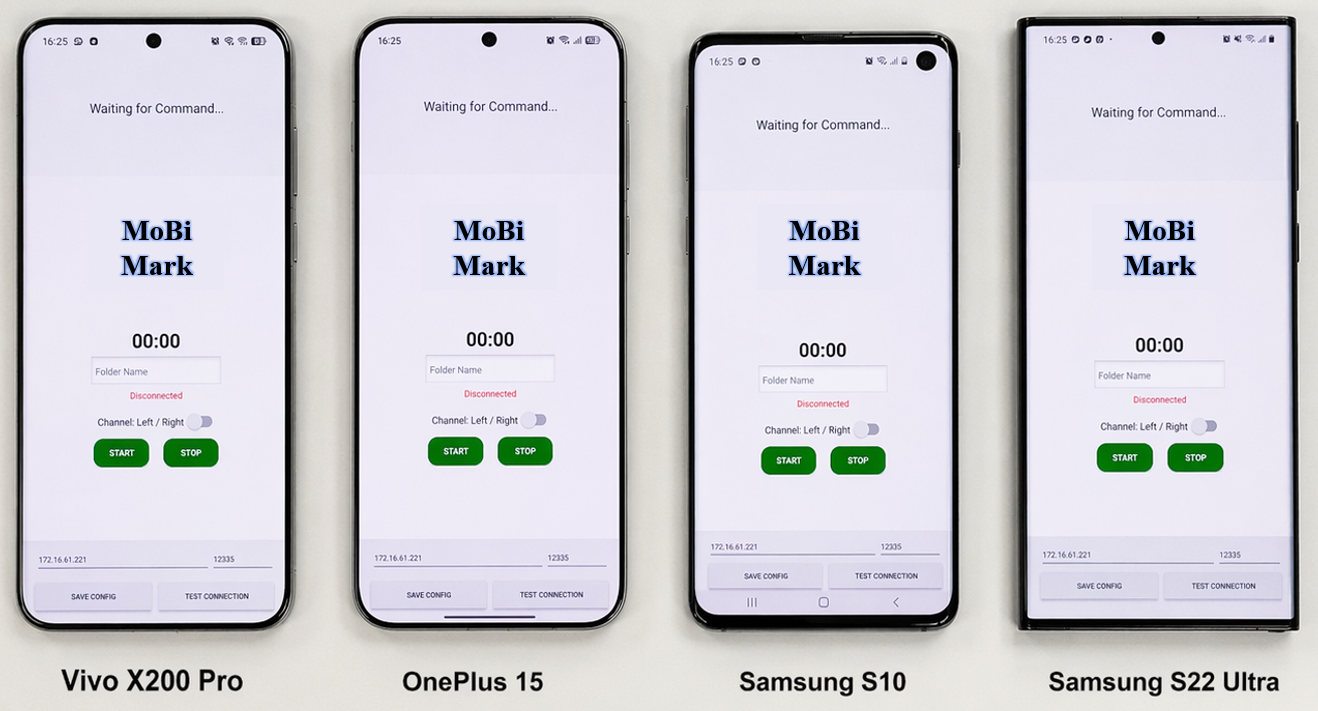}
  \caption{Data capture system and different smartphone devices.}
  \label{fig:dataset}
  \vspace{-0.15in}
\end{figure}

\subsubsection{Speech-ultrasound Acquisition System}
\begin{figure}[t]
  \centering
  \includegraphics[width=0.7\linewidth]{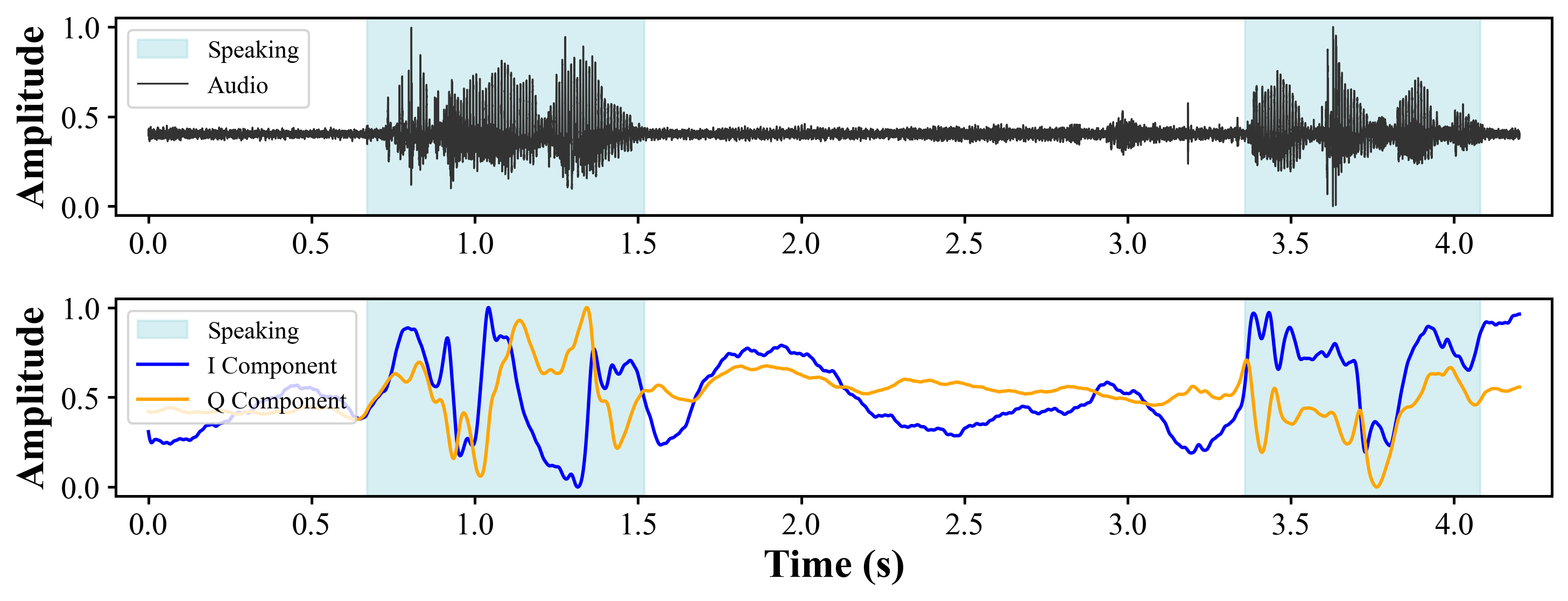}
  \caption{Speech and ultrasound signal synchronization.}
  \label{fig:sync}
  \vspace{-0.2in}
\end{figure}

We constructed a synchronized speech-ultrasound acquisition system using MATLAB and Android Studio, deployed on laptops and smartphones, respectively, as shown in Fig.~\ref{fig:dataset}. 
We use the built-in top earpiece to transmit ultrasound and the adjacent microphone to record audio at 48 kHz, enabling stable near-field acoustic sensing~\cite{zhou2021robust}.
We transmit a Continuous Wave (CW) signal $S(t)$ as a superposition of $N=16$ orthogonal sub-carriers: $S(t) = \sum_{k=0}^{N-1} \cos(2\pi f_k t)$, with $f_k$ ranging from 17.15 kHz to 22.4 kHz at 350 Hz intervals.
Non-rigid articulatory movements perturb the acoustic paths and encode kinematic information into the received signal $R(t)$, which is the spatial integral of scattering echoes over the facial surface~\cite{wang2016device}.
To recover baseband motion information, we perform digital coherent demodulation on $R(t)$ using the locally generated in-phase carrier $\cos(2\pi f_k t)$ and quadrature carrier $-\sin(2\pi f_k t)$ for each $f_k$.
We then apply a 50 Hz low-pass filter to extract baseband articulatory components (speech organ movement frequency is typically around 40 Hz~\cite{lu2018lippass}) and downsample the signals by $D=16$ to 3 kHz.
This process yields complex baseband signals $B_k(t) = I_k(t) + jQ_k(t)$, where $I_k(t)$ and $Q_k(t)$ capture the instantaneous motion-induced in-phase and quadrature components.
The 32 resulting channels (16 frequencies $\times$ I/Q) form the ultrasonic sample.
In parallel, we extract audible speech from the recorded signal using a Butterworth low-pass filter of order six with a cutoff frequency of 16 kHz.
Because both modalities are derived from the same recorded signal, they share the same acquisition timeline and are naturally aligned in time, aligned with speech via hardware timestamps (48 kHz sampling, every 512-sample block), as illustrated in Fig.~\ref{fig:sync}.

\subsubsection{Dataset Setup}
%
Data were collected from 50 users (25 male and 25 female), with 40 randomly selected users for training and the remaining 10 users for testing. Each user contributed approximately 0.5 hours of paired speech and ultrasonic data. During collection, subjects held the smartphone 20--25 cm from the face, approximating a real voice-call scenario. Speech was sampled at 16 kHz and segmented into 5 s windows synchronized with the ultrasonic signal.
The dataset consists of positive and negative samples:

(1) Positive samples (Pos., about 25 h): Genuine synchronized speech-ultrasound pairs collected from legitimate users.

(2) Negative samples in source-level attacks: 
\textbf{Forged Speech (S1, 25 h)}, AI-generated or replayed target speech without any corresponding ultrasonic signal.
\textbf{Forged Speech + Fake Ultrasound (S2, about 250 h)}, forged target speech accompanied by artificially generated ultrasonic signals unrelated to the spoken content.
\textbf{Adaptive Speech-Ultrasound Attack (S3, over 2 h)}, attacker imitates the target user's articulation while replaying forged speech to generate synchronized ultrasonic echoes.

(3) Negative samples in propagation-level attacks:
\textbf{Speech Editing (P1, over 2 h)}, randomly selected 0.5--1.0 s segment of an authenticated speech recording is replaced using a voice editing model while preserving the original watermark.
\textbf{Speech Cropping (P2, over 2 h)}, randomly selected 0.5--1.0 s segment is removed from an authenticated speech recording while retaining the original watermark.
\textbf{Speech Splicing (P3, over 3.5 h)}, 2--5 speech segments from different recordings are concatenated to create forged content, while watermark information extracted from another authenticated recording is injected into the forged speech.
\textbf{Watermark Reuse (P4, over 2 h)}, watermark information extracted from an authenticated recording, including the correct UID and SID, is directly injected into other speech samples without modifying the speech content.
\textbf{Voice Forgery (P5, over 3.5 h)}, voice cloning or editing model synthesizes arbitrary utterances from a few seconds of the target user's speech, after which watermark information extracted from another authenticated recording is embedded into the forged speech.

For the Stabilizer module, a 20-hour clean dataset is constructed from the collected speech, along with corresponding samples subjected to different benign modifications, including adding different environment noises, the noise level is within 5-20\% of the original speech intensity. The training dataset and the test dataset are divided in a ratio of 8:2.
All experiments received our institutional review board (IRB) ethical approval before the start of the study.

\subsubsection{Training Details}
All models are trained on a single NVIDIA RTX 5070 Ti GPU.
For the watermark generation module, Stage 1 trains the ultrasound encoder for 400 epochs with batch size 32, AdamW optimizer ($lr=3\times10^{-4}$, weight decay $1\times10^{-4}$), and cosine annealing ($\eta_{\min}=10^{-6}$, $T_{\max}=400$).
Stage 2 trains the fusion module for 400 epochs with batch size 128, AdamW ($lr=5\times10^{-4}$, weight decay $1\times10^{-4}$), $\alpha=1.0$, with Stage 1 weights frozen; learning rate ($lr$) follows cosine annealing ($\eta_{\min}=10^{-6}$, $T_{\max}=40$) adjusted every 10 epochs.
The Embedder and Detector jointly fine-tune the open-source AudioSeal model using a group-based watermark framing strategy.
The Stabilizer trains with AdamW ($lr=2\times10^{-4}$, weight decay $1\times10^{-4}$, batch size 24, 400 epochs), with Automatic Mixed Precision and gradient clipping (max norm 5.0).
%
For QAT, we fine-tune the pre-trained FP32 model using AdamW with an initial learning rate of $1\times10^{-5}$, weight decay of $1\times10^{-5}$, cosine annealing, a batch size of 64, and 200 epochs. We adopt the QNNPACK backend for mobile-oriented INT8 quantization, while keeping the same loss function as in pre-training.

\subsubsection{Evaluation Metrics}

The performance of the proposed Watermark Generator and Detector modules is rigorously evaluated on the aforementioned test dataset using standard classification metrics. 
The evaluation employs True Positive Rate (TPR), False Positive Rate (FPR), True Negative Rate (TNR), and False Negative Rate (FNR) as primary measures. 
Additionally, the Equal Error Rate (EER), which is the operating point at which FPR equals FNR, is reported to quantify the inherent trade-off between security and usability. 
The assessment is further complemented by the Area Under the Receiver Operating Characteristic (ROC) Curve (AUC), computed from TPR and FPR, and overall classification accuracy, defined as the proportion of correctly classified instances in the corresponding test dataset.
For the embedding and detection module, we follow the evaluation methodology of AudioSeal, employing Scale-Invariant Signal-to-Noise Ratio (SI-SNR) \cite{luo2018tasnet}, Perceptual Evaluation of Speech Quality (PESQ) \cite{rix2001perceptual}, Virtual Speech Quality Objective Listener (ViSQOL) \cite{hines2012visqol}, and Short-Time Objective Intelligibility (STOI) \cite{taal2010short} to assess the quality of the watermarked speech. These are objective perceptual metrics for measuring speech signal quality.

\begin{figure}[t]
  \centering
  \includegraphics[width=0.8\linewidth]{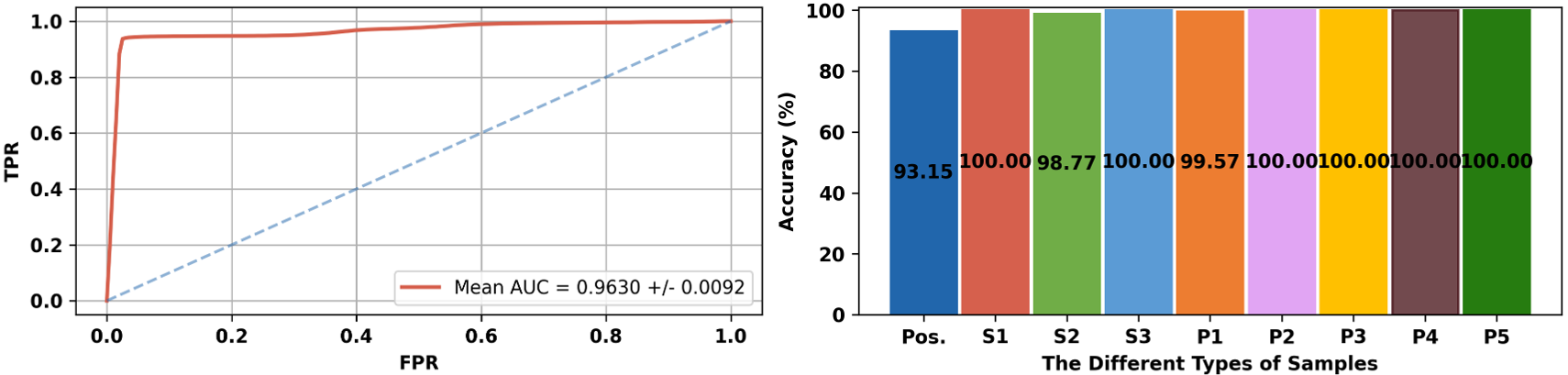}
  \caption{The ROC curve with AUC (left) and accuracy of different types of samples (right).}
  \label{fig:evaluation_results}
\end{figure}

\subsection{Overall Performance}
We randomly select enrollment samples 20 times to compute the average ROC curve.
As shown in Fig.~\ref{fig:evaluation_results}, overall AUC = 0.9630 $\pm$ 0.0092, EER = 0.0558, and positive input accuracy reach 93.15\%.
Although quantization has been applied, the overall accuracy remains at a high level.
These results provide a solid foundation for practical deployment.

\subsubsection{Evaluation of Source-level Attacks}

The evaluation results for source-level attacks are shown in Fig.~\ref{fig:evaluation_results}. The proposed system achieves consistently high detection performance, with an average defense accuracy of 99.59\% across the three attack scenarios. The first two attacks are effectively rejected because forged speech cannot establish consistent cross-modal correspondence with the accompanying ultrasonic signals. For the adaptive attack, although the attacker attempts to synchronize speech with imitated articulatory motions, the resulting ultrasonic echo patterns remain fundamentally different from those of genuine users. Consequently, the proposed multimodal consistency verification generalizes well to all source-level attacks while maintaining a low false rejection rate for legitimate users.

\subsubsection{Evaluation of Propagation-level Attacks}

The defense results against propagation-level attacks are shown in Fig.~\ref{fig:evaluation_results}. Overall, the system achieves an average detection accuracy of 99.91\% across all attack scenarios. Since authentication requires consistency between the speech content and the embedded watermark, semantic modifications inevitably introduce detectable watermark mismatches. The most challenging attacks are P1 and P2, where only a bit of segment is edited or removed while the attacker attempts to preserve the original watermark. Even under this strong assumption, the average similarity of the watermark decreases to 76.59\% ± 9.84\% and 81.02\% ± 8.95\%, respectively, compared to 99.73\% ± 0.84\% for genuine samples. For P3--P5, splicing, watermark reuse, and speech forgery further introduce substantial inconsistencies between the embedded watermark and speech semantics, allowing the proposed method to reliably detect all propagation-level attacks.




\subsubsection{Embedding and Detection Performance}


We evaluate the embedding module using the speech-quality metrics described above, and the results are summarized in Table~\ref{tab:quality}. The results show that the embedded watermark preserves both perceptual quality and speech recognition performance. Moreover, INT8 quantization introduces only negligible performance degradation compared with FP32.
Following the evaluation protocol of AudioSeal, we assess the robustness of the Detector under both benign modifications (e.g., resampling, reencoding, and compression) and malicious attacks (e.g., simple deletion, simple splicing, silencing). As shown in Table~\ref{tab:detector}, FP32 and INT8 models achieve comparable performance, demonstrating the effectiveness of INT8 deployment on edge devices.

\begin{table}[h]
  \centering
  \caption{Speech quality metrics at different quantization levels.}
  \begin{tabular}{ccccc}
    \toprule
    Level & SI-SNR$\uparrow$ (dB) & PESQ$\uparrow$ & STOI$\uparrow$ & ViSQOL$\uparrow$ \\
    \midrule
    FP32 & 23.88 & 4.43 & 0.997 & 4.81 \\
    INT8 & 20.01 & 4.26 & 0.981 & 4.67 \\
    \bottomrule
  \end{tabular}
\label{tab:quality}
\end{table}

\begin{table}[h]
  \centering
  \caption{Detection results (FP32 / INT8) under benign (top) and malicious (bottom) operations.}
  \scalebox{0.85}{
  \begin{tabular}{c cccc}
    \toprule
    Operation & TPR & FPR & TNR & FNR \\
    \midrule
    Compression      & 1.00/1.00 & 0.03/0.03 & -- & -- \\
    Reencoding       & 1.00/1.00 & 0.02/0.03 & -- & -- \\
    Resampling       & 0.98/0.98 & 0.01/0.01 & -- & -- \\
    Noise suppression& 0.98/0.98 & 0.01/0.01 & -- & -- \\
    \midrule
    Simple deletion  & -- & -- & 0.99/0.99 & 0.01/0.01 \\
    Simple splicing  & -- & -- & 0.99/0.99 & 0.09/0.09 \\
    Silencing        & -- & -- & 0.98/0.97 & 0.02/0.02 \\
    Substitution     & -- & -- & 0.99/0.99 & 0.00/0.00 \\
    Reordering       & -- & -- & 0.98/0.97 & 0.00/0.00 \\
    Text-to-speech   & -- & -- & 1.00/1.00 & 0.00/0.00 \\
    Voice conversion & -- & -- & 0.98/0.98 & 0.01/0.01 \\
    \bottomrule
  \end{tabular}}
\label{tab:detector}
\vspace{-0.1in}
\end{table}

\subsubsection{Stabilizer Performance}

\begin{figure*}[t]
    \centering

    \begin{subfigure}[t]{0.30\textwidth}
        \centering
        \includegraphics[width=\linewidth]{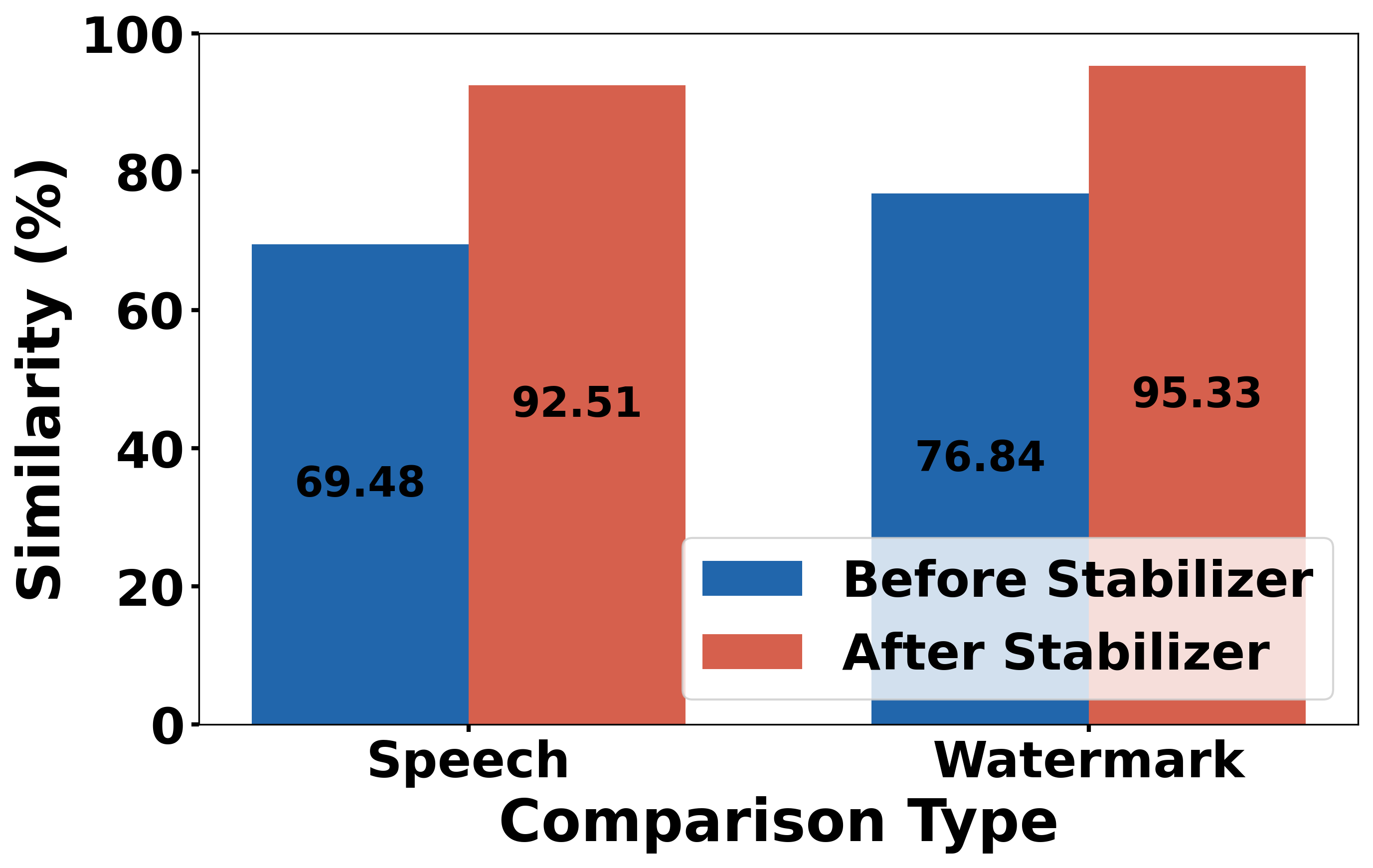}
        \caption{Stabilizer effect.}
        \label{fig:impact_stabilizer}
    \end{subfigure}
    \hfill
    \begin{subfigure}[t]{0.30\textwidth}
        \centering
        \includegraphics[width=\linewidth]{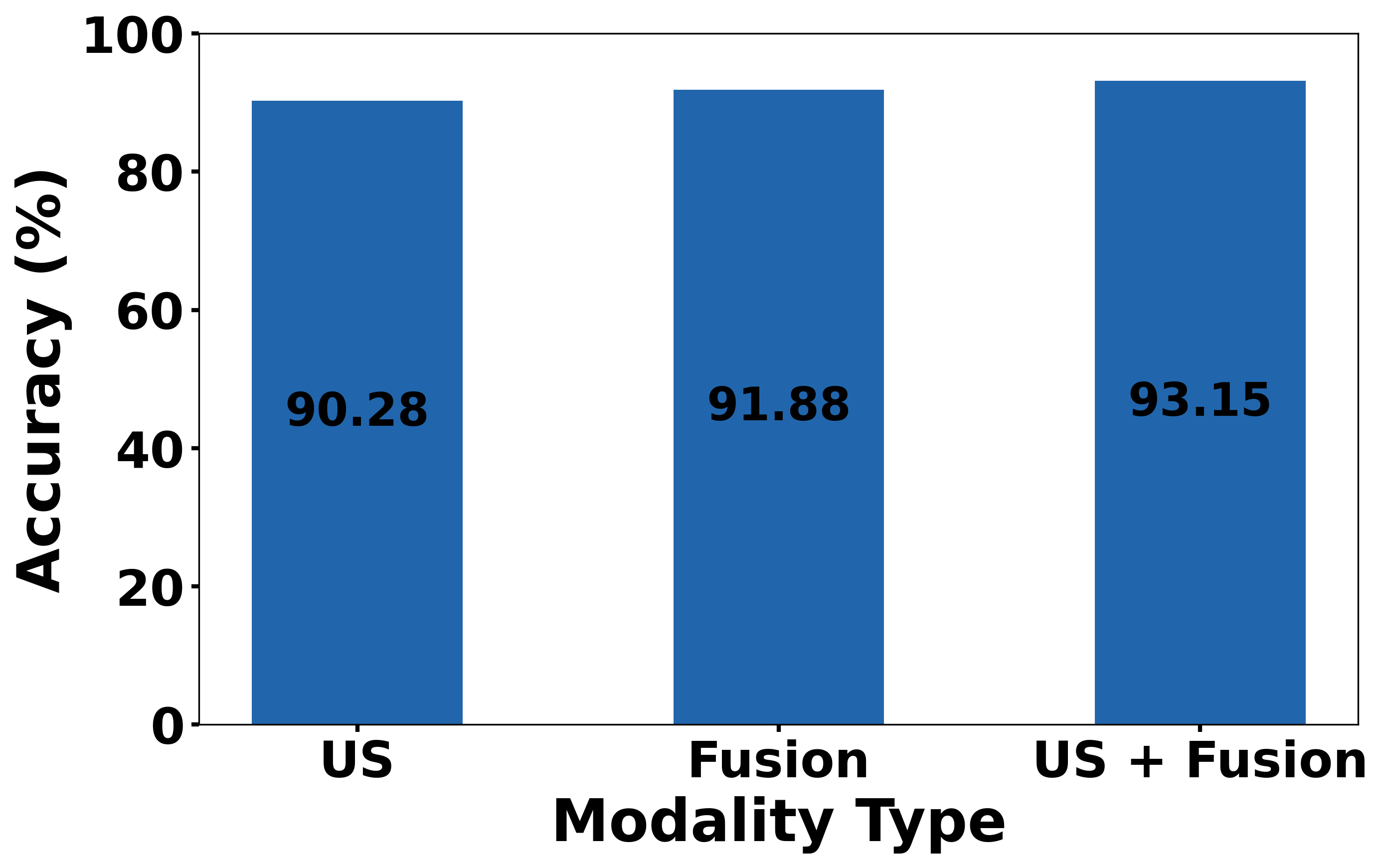}
        \caption{Modality comparison.}
        \label{fig:impact_modality}
    \end{subfigure}
    \hfill
    \begin{subfigure}[t]{0.30\textwidth}
        \centering
        \includegraphics[width=\linewidth]{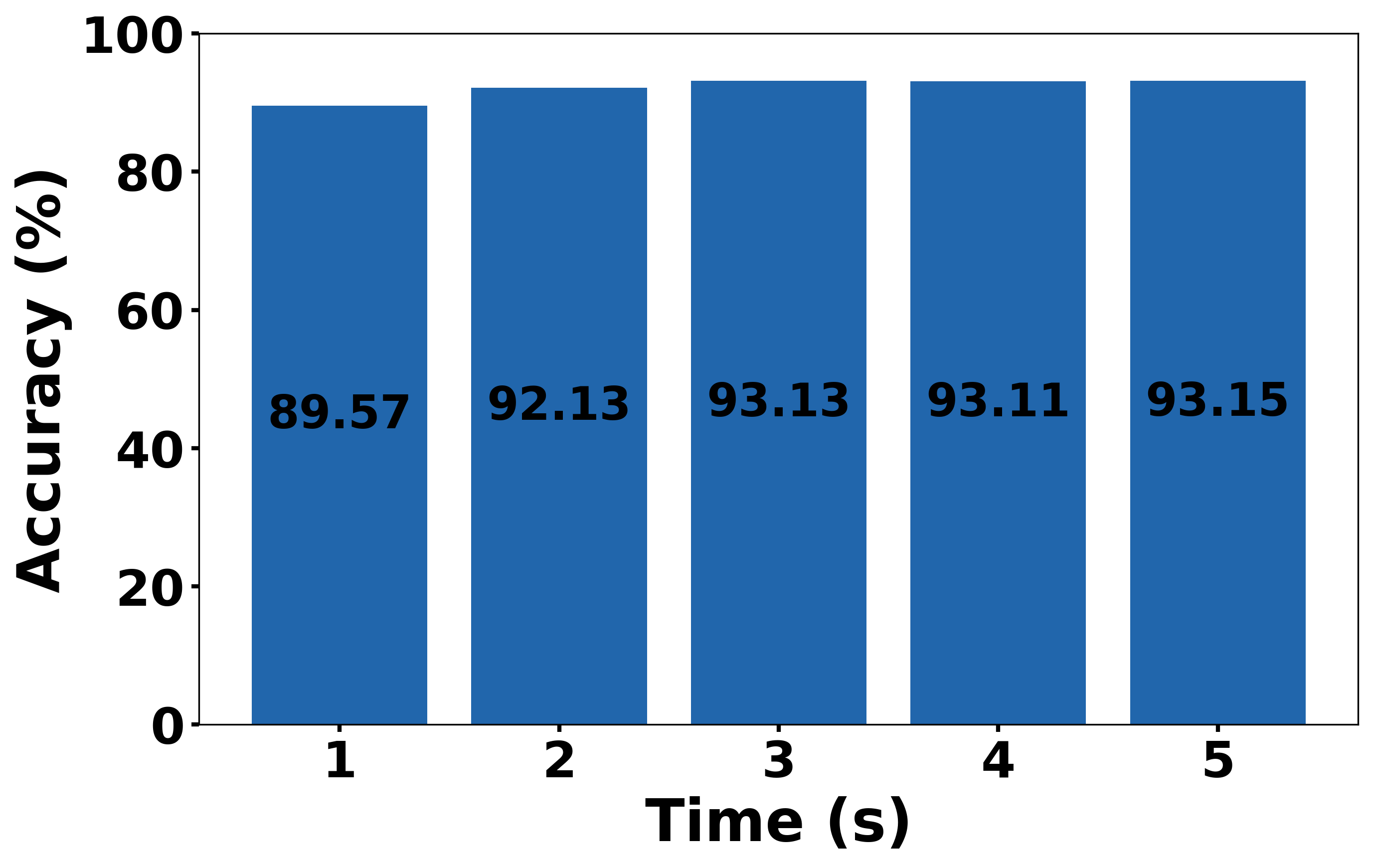}
        \caption{Time window.}
        \label{fig:impact_time}
    \end{subfigure}

    \vspace{0.1in}

    \begin{subfigure}[t]{0.30\textwidth}
        \centering
        \includegraphics[width=\linewidth]{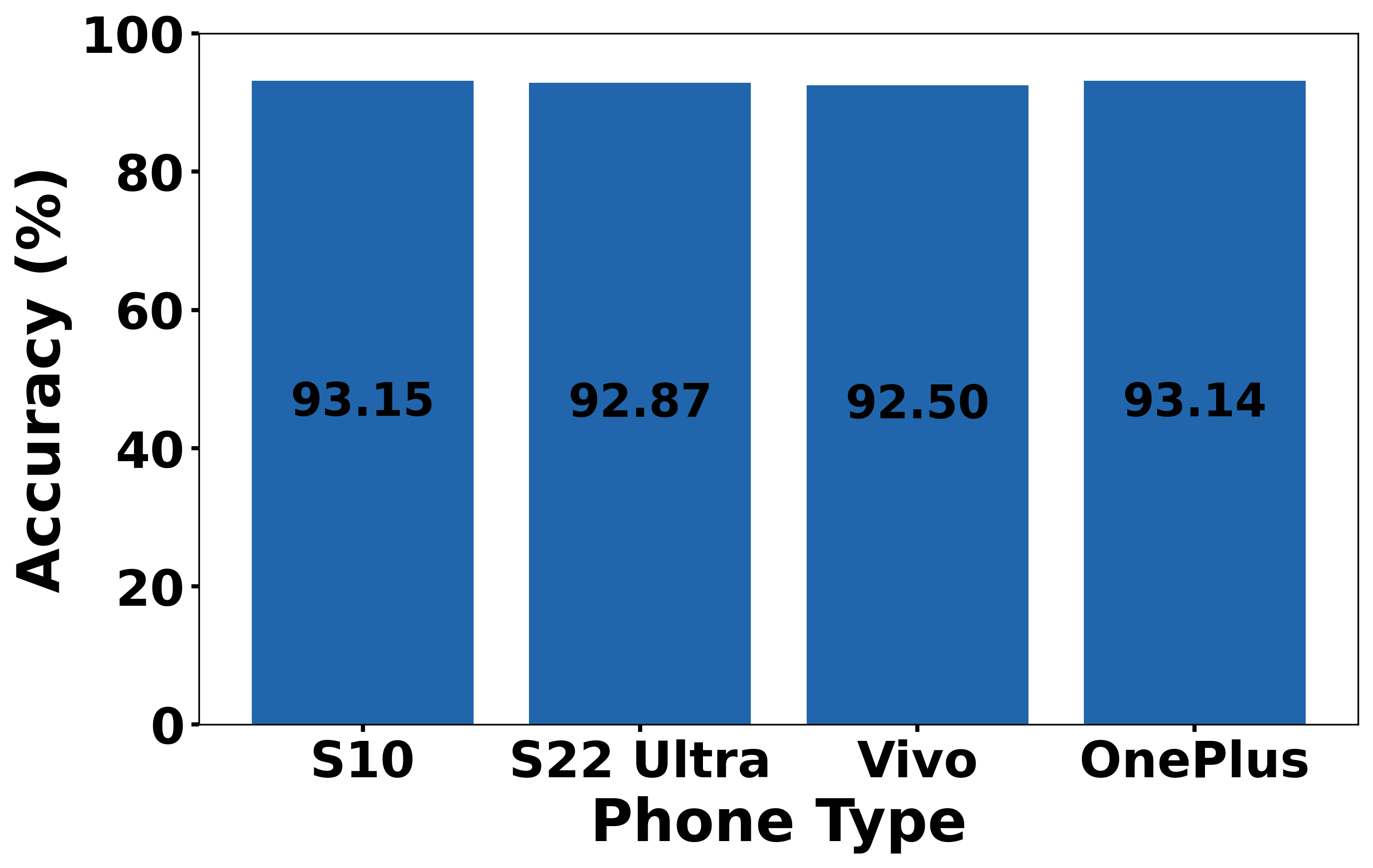}
        \caption{Phone type.}
        \label{fig:impact_phone}
    \end{subfigure}
    \hfill
    \begin{subfigure}[t]{0.30\textwidth}
        \centering
        \includegraphics[width=\linewidth]{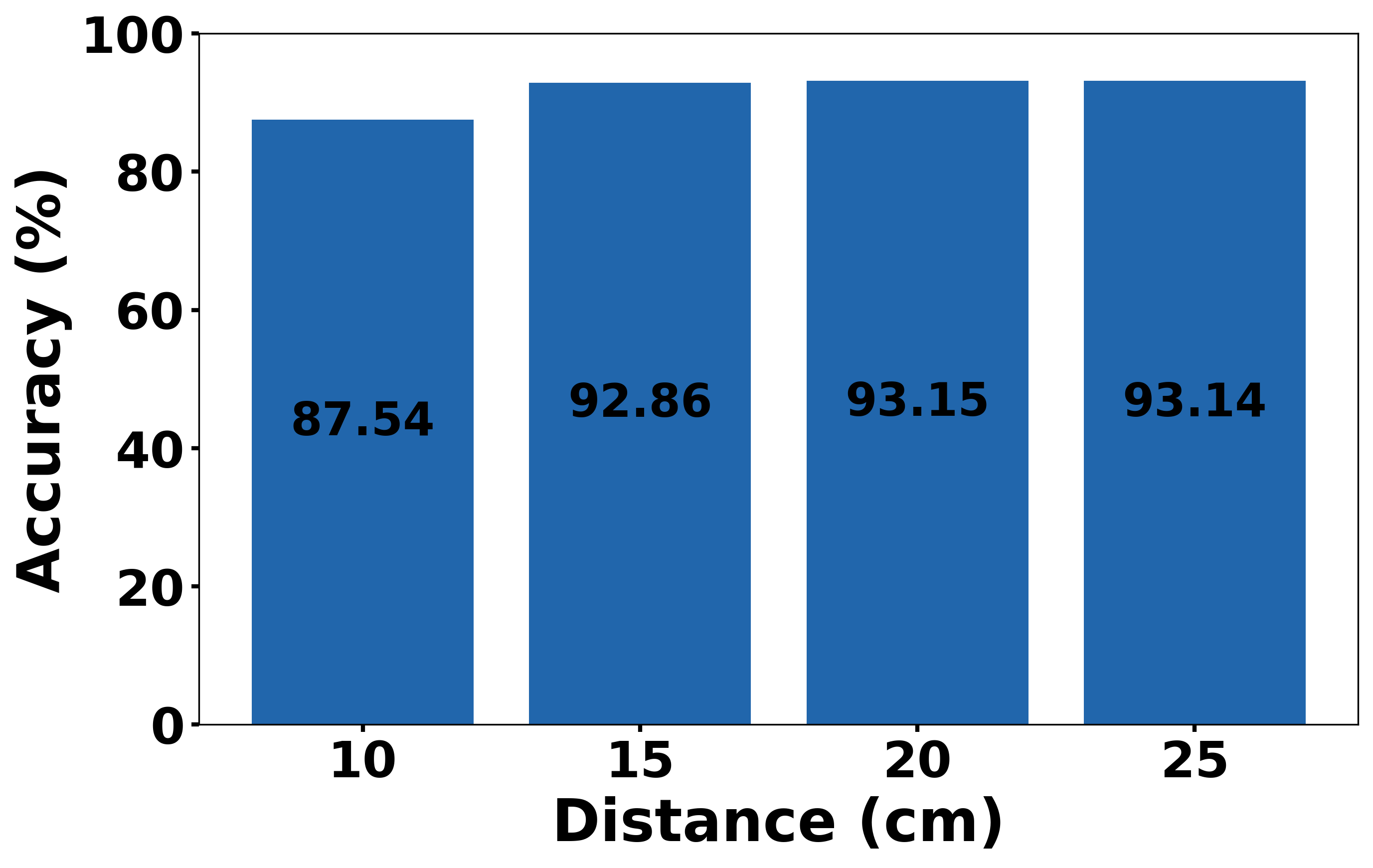}
        \caption{Distance.}
        \label{fig:impact_distance}
    \end{subfigure}
    \hfill
    \begin{subfigure}[t]{0.30\textwidth}
        \centering
        \includegraphics[width=\linewidth]{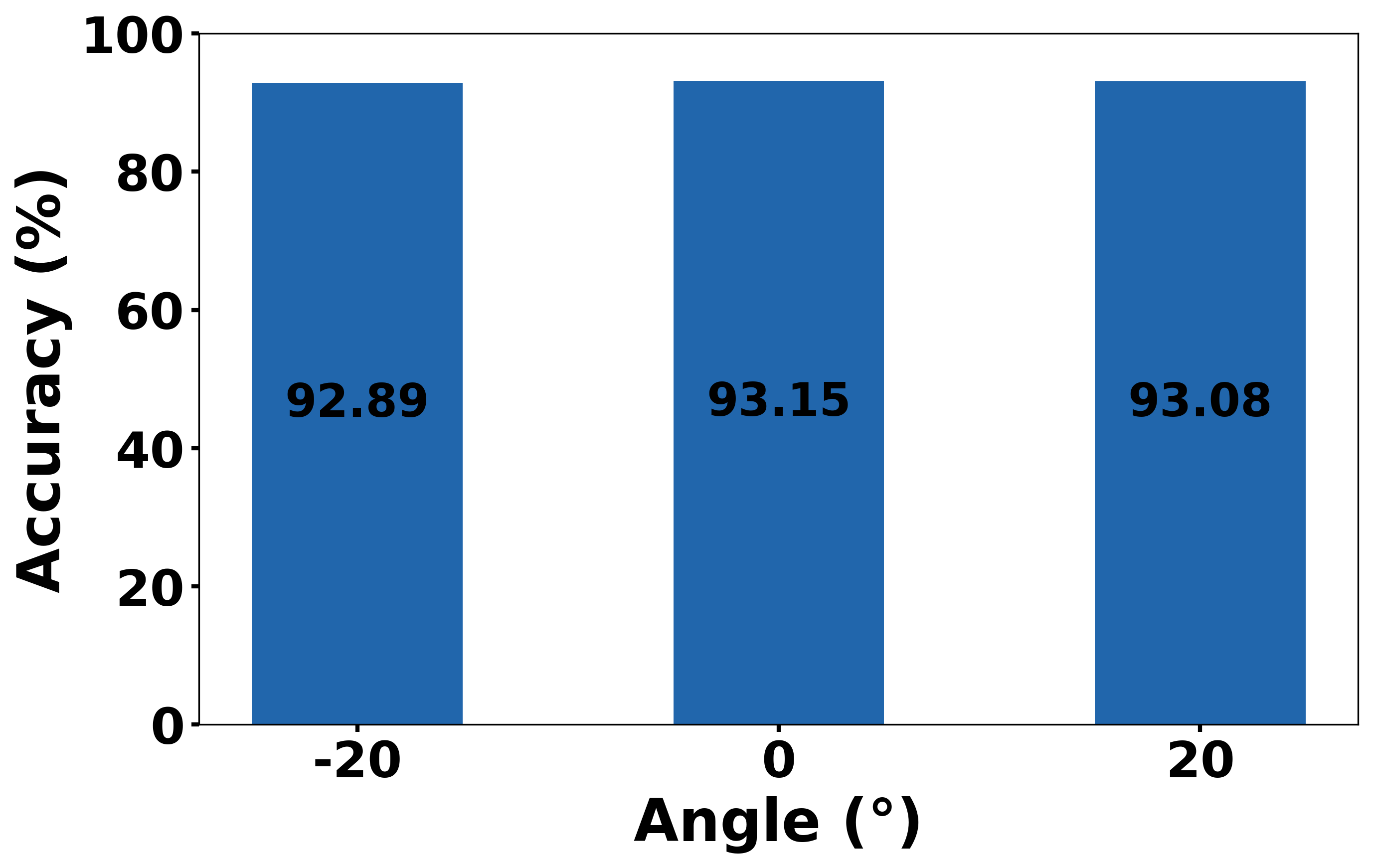}
        \caption{Angle.}
        \label{fig:impact_angle}
    \end{subfigure}

    \caption{Impact of different factors: (a) stabilizer; (b) modality; (c) time window; (d) phone type; (e) distance; (f) angle.}
    \label{fig:impact_all}
    \vspace{-0.2in}
\end{figure*}

For the Stabilizer, there are both quantized and non-quantized versions. At the edge‑side system, we can selectively use the quantized version for source speech processing, while on the server side, quantization is not required. 
We compare the results using the quantized stabilizer to verify its capability; the performance of the non‑quantized stabilizer is no worse than the quantized one. 
Based on metrics from the detection side, we use the test speech samples containing original clean utterances and benign operation (noisy samples) for testing the Stabilizer we train. 
These noisy samples are fed into the stabilizer, and we compare the cosine similarity between the features of the stabilizer‑processed speech and the original clean speech, as well as the similarity between the extracted watermark information (after passing through our system) and the original watermark. The results are shown in Fig. \ref{fig:impact_stabilizer}.
It can be seen that, for both the speech itself and the watermark information, the accuracy improves by 33.15\% and 24.06\%, respectively, after the stabilizer is applied, the watermark code similarity reaches 95.33\%, which demonstrates the effectiveness of adding the Stabilizer.

\subsection{Evaluation of Impact Factors}


For the various influencing factors considered in this experiment, which pertain solely to the discrimination of source input. We evaluate all factors using the positive samples accuracy metric of the Watermark Generator.
It is worth noting that, unlike software-based watermarking schemes, ultrasonic watermark generation is intentionally constrained by the physical characteristics of acoustic propagation. Successful embedding therefore requires the user's active participation and an effective distance and orientation between the smartphone and the speaker.
%
Authentication systems generally favor stricter decision boundaries, as the cost of false acceptance outweighs that of false rejection.


\subsubsection{Impact of Different Features}
In the decision logic for target recognition, we employ a method that collaboratively discriminates using ultrasonic signals and fused feature vectors. To investigate the contribution of both to the final recognition performance, this section quantifies the influence of ultrasonic and fused vectors, with the results shown in Fig. \ref{fig:impact_modality}. The experiments demonstrate that the system achieves the highest recognition accuracy 93.15\% when the two modalities are used jointly. This result confirms that the trained fused feature vector effectively captures the deep-level correspondence between the speech and ultrasonic modalities, rather than performing simple feature concatenation, while ensuring that speech information does not interfere with the discriminative ability of the ultrasonic modality, thereby reflecting the effectiveness of the multimodal fusion design in the proposed model.


\subsubsection{Impact of Different Time Windows}\label{sec: different time window}
The time window length (T) is a critical parameter in dynamic speech watermarking systems. Its value not only influences the discrimination accuracy of ultrasonic signals but also directly determines the computational complexity of the watermark generation model. A larger T corresponds to a longer watermark update cycle, which is beneficial for reducing computational overhead. Conversely, a smaller T, while capable of reducing system latency, increases the computational load for watermark generation and exerts a certain impact on the discrimination accuracy of ultrasonic signals. This section verifies the system’s tunability between latency and performance by evaluating the change in user identification accuracy under different T settings, as shown in Fig. \ref{fig:impact_time}. As T increases, the accuracy exhibits a gradual upward trend. Even under a short window of T=1s, the system can maintain an identification accuracy of approximately 89.57\%, indicating that the proposed mechanism can achieve an effective balance between latency and recognition performance through flexible configuration of T. 
It is notable that all of our models are constructed using sequence-length-invariant architectures, allowing users to freely adjust the temporal window size without the need for retraining, which facilitates greater generalizability in practical deployment scenarios.


\subsubsection{Impact of Different Smartphone Types}
To verify the generalization capability of the system across different mobile devices, this study selects multiple phone models, including Samsung S10 (S10), Samsung S22 Ultra (S22 Ultra), vivo X200 Pro (Vivo), and OnePlus 15 (OnePlus), for universal applicability testing. The experimental results, shown in Fig. \ref{fig:impact_phone}, demonstrate that the system maintains stable identity authentication performance on all devices, with an accuracy of no less than 92.50\%. This is primarily attributed to the fact that the ultrasonic signals rely only on the microphone hardware commonly integrated in smartphones, without the need for specialized sensors, thereby ensuring good device compatibility and deployment flexibility. The system is suitable for mobile terminals across different tiers, providing a feasible solution for privacy protection in practical voice interaction scenarios.


\subsubsection{Impact of Different Distances}\label{sec: different distance}
The relative distance between the user and the device is one of the key factors affecting the quality of ultrasonic signals. To evaluate the impact of distance variation on system robustness, we conduct tests within the range of 10 cm to 25 cm, with 5 cm intervals, simulating typical voice interaction distances. The results, shown in Fig. \ref{fig:impact_distance}, demonstrate that the system exhibits stable recognition performance across different distances, with an overall accuracy maintained above 87.54\%, an average 91.67\%, indicating that the proposed method possesses strong adaptability to distance changes.


\subsubsection{Impact of Different Angles}\label{sec: different angle}
Furthermore, considering that in actual usage scenarios, the phone screen and the user’s face often exhibit a certain angle, we investigate the impact of different relative angles on system performance. Taking the parallel alignment of the screen and the facial plane as the 0° baseline, data are collected for device rotation within ±20° in the vertical direction, an angular range that ensures normal transmission and reception of ultrasonic signals by the microphone. The experimental results, shown in Fig. \ref{fig:impact_angle}, indicate that the system maintains high recognition robustness under various angular conditions, achieving an accuracy of no less than 92.89\%, which demonstrates that the proposed method exhibits good fault tolerance for posture variations.


\subsection{System Inference and Privacy}

\begin{figure}[t]
    \centering
    \includegraphics[width=0.3\linewidth]{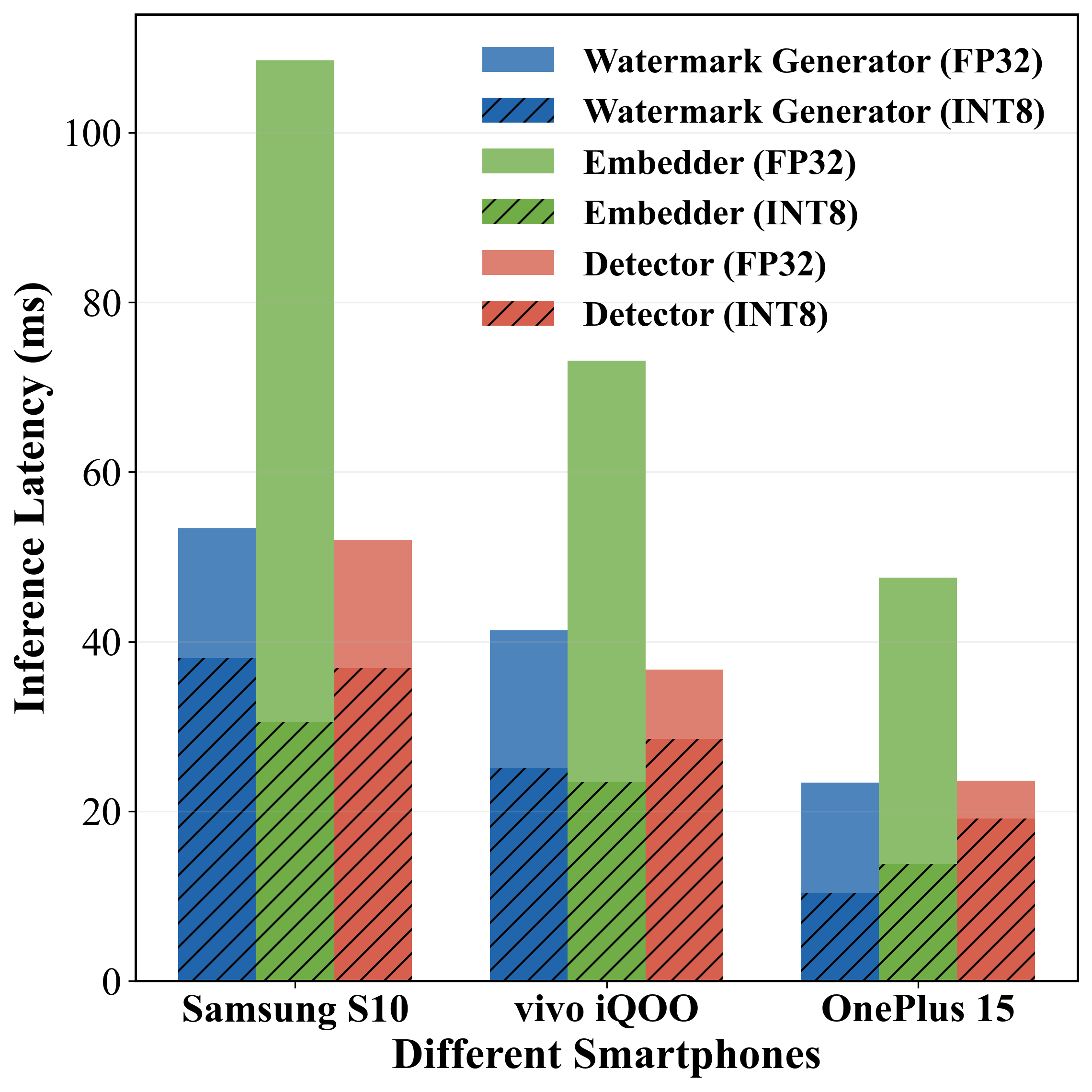}\hfill
    \includegraphics[width=0.3\linewidth]{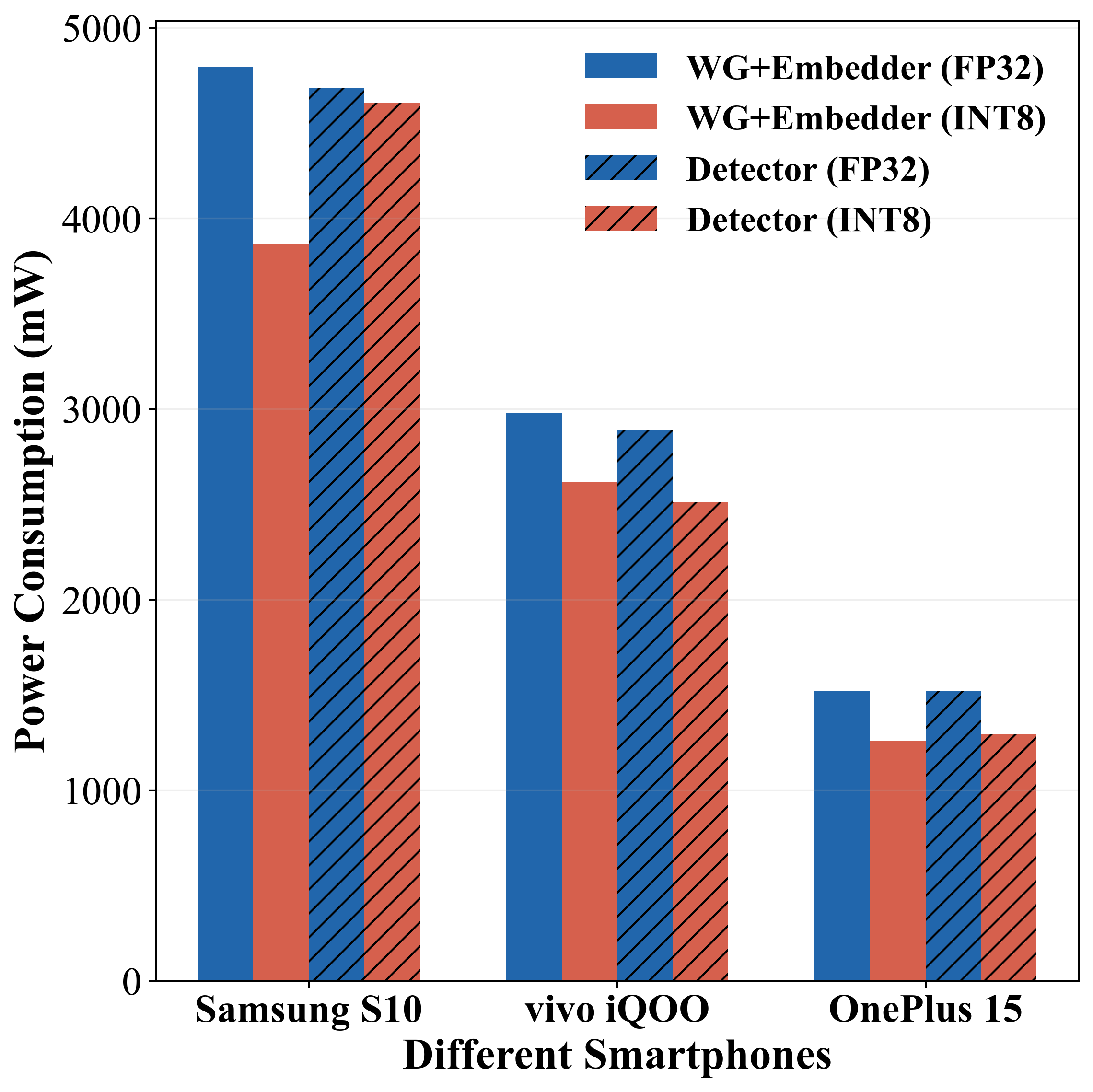}\hfill
    \includegraphics[width=0.3\linewidth]{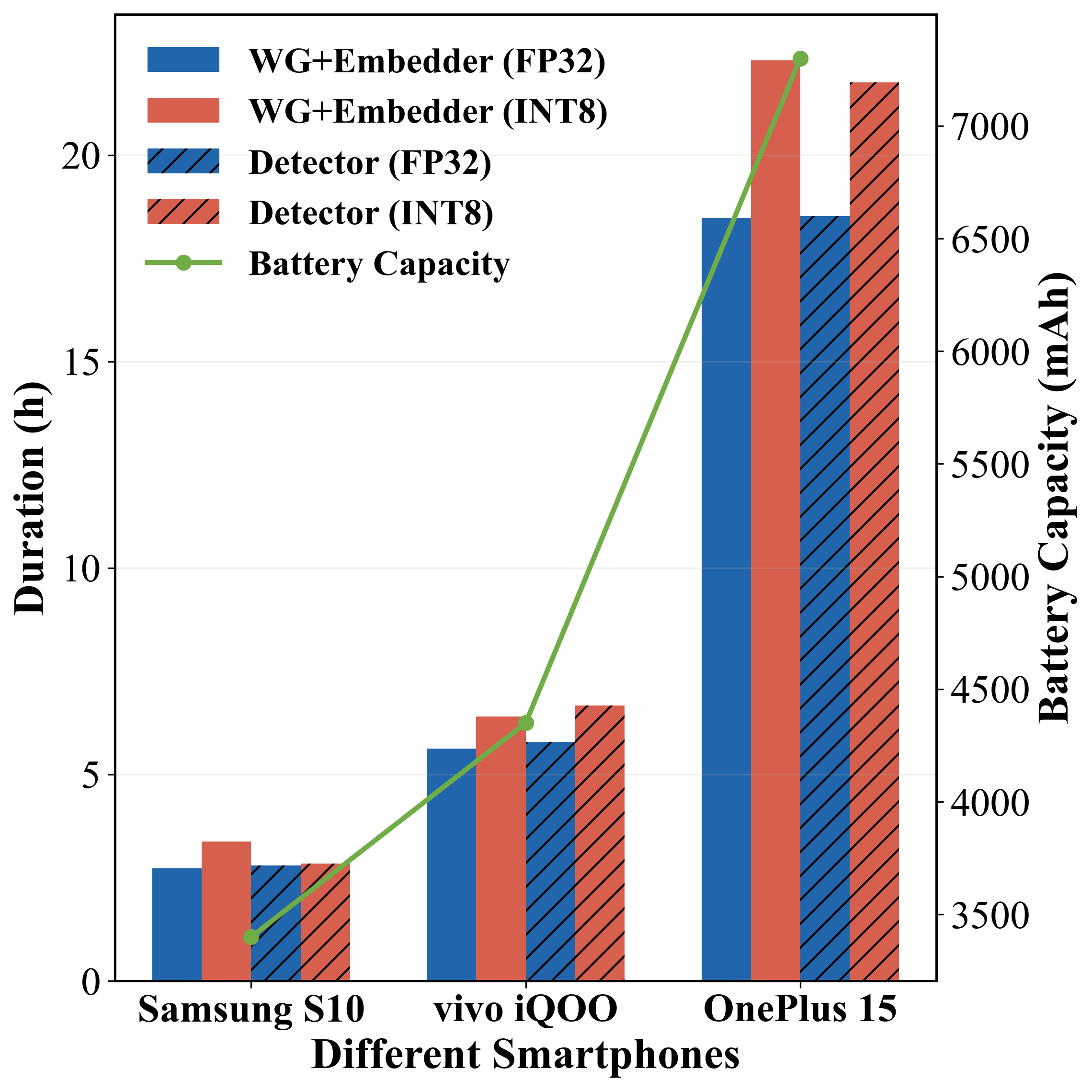}
    \caption{Inference latency (left), power consumption (center), and battery duration (right) on different smartphones.}
    \label{fig:device_perf}
    \vspace{-0.2in}
\end{figure}

Our system is deployed on three types of smartphones: Samsung S10, vivo iQOO8, and OnePlus 15, released in 2019, 2021, and 2025 respectively, providing a diverse baseline for evaluating system robustness.
The testing scenario simulates a real-time voice interaction.
For the Watermark Generator, a sufficiently long temporal window is required to generate the watermark; a window length of 1 s is used in the test.
For the Embedder and Detector, a sliding step of 20 ms is set to balance latency and real-time requirements, meaning every 20 ms the system generates a watermark based on the preceding 1 s of audio and embeds it into the current window.
A total of 100 samples verifies the real-time capability.

The latency results are shown in Fig.~\ref{fig:device_perf}.
On our INT8-quantized implementation, the Watermark Generator and Embedder modules achieve minimum total latency of $10.32 + 13.80 = 24.12$ ms, meeting the demands of real-time communication~\cite{series2003transmission}.
INT8 quantization achieves latency reduction of up to 71.9\% and power-consumption reduction of up to 19.35\% compared to FP32.
Newer smartphones, due to their better support for on-device model deployment, exhibit lower power consumption and longer sustained operation---up to 18.47 hours on the latest device.
These results fully validate the real-time online performance and the effectiveness of model lightweighting for on-device deployment.
Since our model is independent of the temporal window, users can adjust the watermark generation and embedding time window according to their hardware capabilities, achieving a trade-off between speed and quality.

Regarding user privacy and security, the entire watermarking and integration process runs entirely on the local side without involving cloud processing, thus ensuring complete privacy protection.
The detection end, which may need cloud access, can be addressed using classic hybrid encryption architecture~\cite{rivest_rsa_1978,nist_aes_2001,ietf_tls13_2018}.
Given the general applicability of our system, it can be effectively deployed on users' mobile edge devices, providing real-time, online protection for voice security.


\section{Discussion}
\label{sec:discussion}

MoBiMark provides an end-to-end framework for speech security by combining real-time biometric source authentication, on-device processing, and dual-layer dynamic watermarking to protect speech against unauthorized editing and forgery during transmission. By leveraging ubiquitous microphones and resource-efficient inference, the system is designed for deployment on commodity smart devices.

MoBiMark relies on ultrasonic sensing to capture user-specific biometric features, which introduces practical constraints on sensing distance and orientation. As the user moves farther from the smartphone, weaker echoes may degrade biometric feature quality and authentication reliability. Rather than being solely a limitation, the effective sensing range also imposes a physical proximity requirement that can strengthen the association between watermark generation and the user's physical presence. Since typical voice interactions occur within a relatively close range of the smartphone, this requirement is compatible with the intended use scenarios. In addition, preprocessing helps suppress motion-induced low-frequency artifacts, improving the reliability of ultrasonic feature extraction under practical conditions.

The security of the system also depends on the integrity of the underlying hardware and sensing pipeline. As future work, we plan to investigate hardware-assisted encryption and data-flow integrity mechanisms to protect internal sensor streams against unauthorized access and malicious data injection. Such mechanisms could further strengthen the system against attacks targeting the trusted acquisition path.

Beyond source authentication and watermark protection, extending MoBiMark to real-time voice communication is a promising direction. Although the current system provides real-time speech processing, speech codecs may distort or invalidate embedded watermarks during transmission. One potential solution is to integrate watermarking directly into neural speech codecs, enabling watermark information to be preserved in latent representations rather than relying solely on compressed waveforms. This could improve watermark integrity across the communication pipeline and facilitate practical deployment in real-time voice calls.

\section{Conclusion}
\label{sec:conclusion}
The proliferation of generative speech models poses critical threats to traditional voiceprint-based real-time voice authentication. 
In this paper, we present MoBiMark, the first real-time online biologic speech watermarking system using smartphones ultrasound signals.
It binds ultrasonic vocal-tract echoes to speech content, embedding a dual-layer dynamic watermark at capture time for source authentication and different attacks detection.
Detailed training and quantization design achieve high accuracy, strong robustness, and efficient on-device operation.
Evaluations on 50 users across four smartphones demonstrate average 99.79\% attack detection accuracy on source and propagation with negligible perceptual impact, 
which enables practical real-time speech security for voice interaction in the future.

\bibliographystyle{ACM-Reference-Format}
\bibliography{sample-base}

\appendix

\end{document}